\documentclass[aps,pra,twocolumn,longbibliography,floatfix,nobibnotes,10pt,superscriptaddress]{revtex4-2}

\usepackage[T1]{fontenc}
\usepackage[utf8]{inputenc}

\usepackage{amsmath,amssymb,amsfonts,bm}
\usepackage{mathtools}
\usepackage{physics}

\usepackage{graphicx}
\usepackage{xcolor}
\usepackage{booktabs}
\usepackage{array}
\usepackage{enumitem}
\usepackage{placeins}
\usepackage{pgfplots}

\pgfplotsset{compat=1.18}

\let\label\KernelLabel
\usepackage[
colorlinks=true,
linkcolor=blue,
citecolor=blue,
urlcolor=blue,
filecolor=blue,
breaklinks=true
]{hyperref}

\makeatletter
\let\ltx@label\label
\makeatother
\usepackage[nameinlink,capitalise,noabbrev]{cleveref}

\graphicspath{{./}}

\newcommand{\affilUKON}{Department of Physics, University of Konstanz, 78457 Konstanz, Germany.}

\makeatletter
\newcommand{\alsolabel}[1]{\ltx@label{{#1}}}
\makeatother

\begin{document}

	\title{Preparation geometry and slow-sector routing in driven Kerr resonators: an operational spectral theory of Liouvillians}
	\author{Kilian Seibold}\affiliation{\affilUKON}

\begin{abstract}
Liouvillian eigenvalues determine decay rates and oscillation frequencies, but not mode excitation, state deformation, or detection. We develop an operational spectral theory based on matched left and right eigenoperators. Left eigenoperators govern excitation; right eigenoperators describe state deformation and readout. The product of excitation and detection factors defines a gauge-invariant modal weight. For coherent bosonic preparations, left eigenoperators define phase-space excitation maps whose zeros locate mode-suppressing preparations. Within a metastable slow sector, matched eigenoperators define operational coordinates and, for stochastic reductions, a transition generator. Slow decay rates and stationary populations fix the transition rates of two-state and symmetric three-state reductions, but in general only constrain them. With an absorbing state, both remaining states can prefer switching to each other only if the faster slow decay rate exceeds three times the slower. In driven Kerr resonators, rotating the phase of a one-photon bias, with compensating detuning and parametric drive, reverses the preferred exit of a metastable state at numerically matched slow decay rates. Given two distinct real slow poles, stationary coordinates, and an invertible readout calibration, two observable means at one delay determine both exit rates of a prepared state. Without further constraints, the same data determine the full slow generator if and only if the preparation excites both slow modes. The shot cost at fixed precision grows asymptotically as the inverse squared distance to a regular excitation-zero contour of the faster slow mode. Preparation geometry thus links relaxation control and identifiability: suppressing the slowest mode can accelerate relaxation, as in the strong Mpemba effect, but suppressing either slow mode prevents full reconstruction from that preparation.
\end{abstract}

\maketitle

\section{Introduction}
\label{sec:intro}

Liouvillian eigenvalues determine decay rates and oscillation frequencies, but not which modes a preparation excites, how they enter an observable, or how metastable populations redistribute. In metastable regimes, a few slow modes govern transitions between long-lived configurations~\cite{Minganti2018,Macieszczak2016,Rose2016,Casteels2017,Macieszczak2021,Brown2024}.

\begin{figure*}[t]
	\centering
	\includegraphics[width=\textwidth]{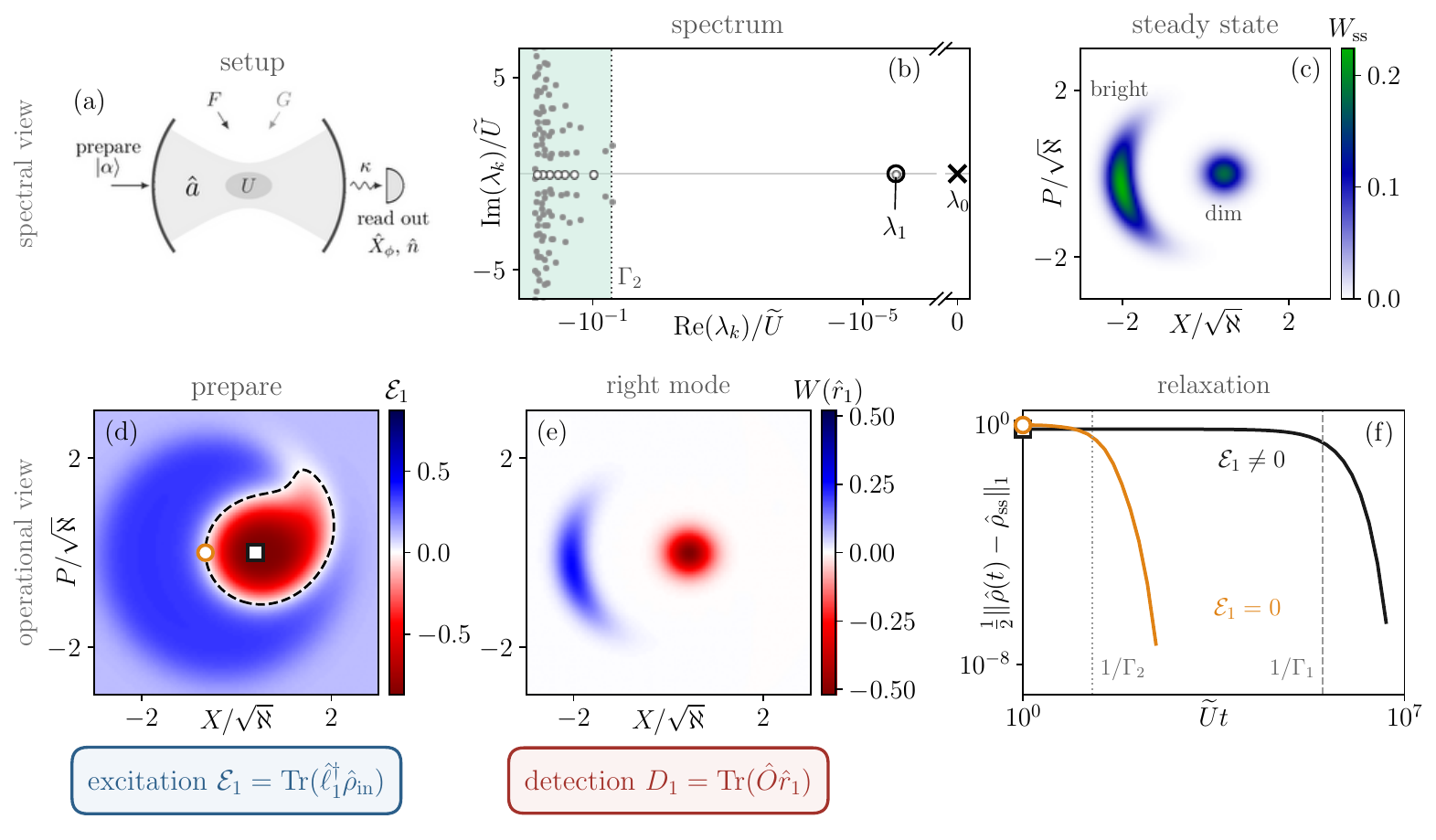}
	\caption{\textit{Spectral and operational views of Liouvillian relaxation.} (a) Driven Kerr resonator with coherent-state preparation $|\alpha\rangle$ and quadrature $\hat X_\phi$ or photon-number $\hat n$ readout. (b) Liouvillian spectrum under linear driving, with real (open) and complex (filled) eigenvalues. The switching mode $\lambda_1=\lambda_{\rm sw}$ is separated from the faster spectrum (shaded), with $\Gamma_2/\Gamma_1=1.7\times10^4$, where $\Gamma_k=-\operatorname{Re}(\lambda_k)$. (c) Stationary Wigner function $W_{\rm ss}(X,P)$ showing the bright and dim configurations. (d) Coherent-state excitation map $\mathcal E_1(X,P)=\langle\alpha|\hat\ell_1^\dagger|\alpha\rangle$. Its zero contour (dashed) identifies preparations with the stationary bright-state coordinate, which need not coincide with a basin boundary. (e) Right-mode Wigner symbol $W_1^R(X,P)$ describing the population contrast between the two configurations, with observable overlap $D_1(\hat O)=\operatorname{Tr}(\hat O\hat r_1)$. (f) Relaxation in trace distance to $\hat\rho_{\rm ss}$ for the preparations marked in (d). When mode 1 dominates, the trace distance is $\tfrac12|\mathcal E_1|\|\hat r_1\|_1e^{-\Gamma_1t}$. The initially more distant preparation suppresses this term and relaxes on the scale $1/\Gamma_2$, whereas the closer preparation relaxes on the scale $1/\Gamma_1$. The sign convention is $\operatorname{Tr}(\hat n\hat r_1)>0$; blue (red) denotes positive (negative) values in (d,e). Parameters: $\Delta/\widetilde U=1.9$, $\widetilde F/\widetilde U=0.5$, $\kappa/\widetilde U=0.1$, $G=0$, and $\aleph=5$ (Sec.~\ref{subsec:kerr_model_protocol}).}
	\label{fig:scheme1}
\end{figure*}

The left-right decomposition~\cite{Briegel1993,Barnett2000,Honda2010,Torres2014,Prosen2010} separates a mode's coupling to an input from its response~\cite{Schomerus2020,Scarlatella2019,Seibold2026}. Large modal amplitudes can make the spectral gap a poor predictor of transient relaxation~\cite{Mori2020,Mori2023,Shirai2024}, and preparations that suppress the slowest mode can accelerate relaxation, as in the strong Mpemba effect~\cite{Klich2019,Carollo2021,Ares2025,Zhang2025,Xia2026}. Open quantum dynamics has been identified from time traces~\cite{Burgarth2014,ZhangSarovar2015,Flammia2026}, and gauge-invariant mode measures have been introduced~\cite{Thomas2026}.

Here we develop an operational spectral theory based on matched left and right eigenoperators. For a resolved mode, the eigenvalue fixes the time dependence, the left eigenoperator the excitation by a preparation, and the right eigenoperator the state deformation and readout overlap. The product of excitation and detection factors, the modal weight, is gauge invariant. Coherent-state overlaps with left eigenoperators define phase-space excitation maps, whose zeros locate preparations that suppress a mode. Right-eigenoperator Wigner symbols describe the state deformations. With a nonzero readout overlap, fitting transients from several preparations to common resolved poles reconstructs these maps up to a factor per mode. In a metastable slow sector, the matched eigenoperators define operational coordinates and, for stochastic reductions, a transition generator.

Driven Kerr resonators offer an experimentally accessible realization of these structures~\cite{Puri2020,Grimm2020,MarthalerDykman2006,Venkatraman2024}. Under linear driving, one switching mode connects a bright and a dim response (Fig.~\ref{fig:scheme1}). Its excitation-zero contour marks the stationary bright-state coordinate, need not coincide with a basin boundary, and shifts with detuning while the right-mode Wigner symbol changes little. Under parametric driving, two opposite-phase lobes coexist with a low-amplitude center, and parity separates the inter-lobe imbalance from population exchange with the center. A one-photon drive breaks parity and biases the competition between flips to the opposite lobe and center capture~\cite{CrosCarrilloDeAlbornoz2024}.

We show that the slow decay rates and stationary populations fix all transition rates in the two-state and parity-symmetric three-state Markov reductions. Without parity, the transition rates are generally underdetermined, but with an absorbing center the slow decay rates $\Gamma_1<\Gamma_2$ still sharply bound the round-trip probability, and both lobes can prefer flipping only if $\Gamma_2>3\Gamma_1$. In the biased resonator, rotating the drive phase while adjusting the detuning and parametric drive reverses a lobe's preferred exit at numerically matched slow decay rates and a nearly unchanged mean residence time. A quantum detection model reproduces the reversal without a three-state reduction.

Given the slow poles, stationary coordinates, and an invertible readout calibration, two observable means at one delay determine both exit rates of a prepared lobe without resolving rare switching events. Without further structural constraints, the same data identify the full generator if and only if the preparation excites both slow modes. Near a regular excitation zero of the faster slow mode, the shot cost grows as the inverse squared distance, if the other information factors stay finite and nonzero. Preparation geometry thus governs both relaxation and identifiability: suppressing the slowest mode can accelerate relaxation, but suppressing either slow mode precludes full reconstruction from that preparation alone.

\section{Preparation and detection of Liouvillian modes}
\label{sec:operational_spectral_theory}

We summarize the standard spectral decomposition of Markovian quantum dynamics and introduce the notation used throughout. We consider a time-independent Gorini--Kossakowski--Sudarshan--Lindblad (GKSL) Liouvillian $\mathcal L$ acting on a finite-dimensional Liouville space, with dynamics $\partial_t\hat\rho(t)=\mathcal L\hat\rho(t)$ and a unique stationary state,
\begin{equation}
	\mathcal L\hat\rho_{\rm ss}=0
	\;,\qquad
	\operatorname{Tr}(\hat\rho_{\rm ss})=1
	\;,\qquad
	\dim(\ker\mathcal L)=1
	\;.
	\label{eq:steady_state}
\end{equation}
The Kerr models studied below satisfy Eq.~\eqref{eq:steady_state} at finite $\aleph$ (Sec.~\ref{subsec:kerr_model_protocol}); their metastability is associated with long-lived nonstationary modes, not with additional stationary states (Appendix~\ref{app:spectral_conventions}). For a diagonalizable Liouvillian, the right and left eigenoperators satisfy~\cite{Briegel1993,Honda2010,Albert2014,Albert2016}
\begin{equation}
	\mathcal L\hat r_k=\lambda_k\hat r_k
	\;,\qquad
	\mathcal L^\dagger\hat\ell_k=\lambda_k^*\hat\ell_k
	\;,\qquad
	\operatorname{Tr}(\hat\ell_k^\dagger\hat r_j)=\delta_{kj}
	\;.
	\label{eq:right_left_eigenoperators}
\end{equation}
Here $\mathcal L^\dagger$ is the Hilbert--Schmidt adjoint and $\lambda_k=-\Gamma_k+i\Omega_k$, with decay rate $\Gamma_k\geq0$ and oscillation frequency $\Omega_k$. The stationary eigenoperators are $\hat r_0=\hat\rho_{\rm ss}$ and $\hat\ell_0=\hat I$. Trace preservation implies $\operatorname{Tr}(\hat r_k)=0$ for $\lambda_k\neq0$, so nonstationary right eigenoperators describe changes in the density matrix, not physical states.

For an input operator $\hat X$ and an observable $\hat O$, we define the excitation factor $E_k$, detection factor $D_k$, and modal weight $w_k$ as
\begin{equation}
	\begin{aligned}
		E_k(\hat X)&=\operatorname{Tr}(\hat\ell_k^\dagger\hat X)
		\;,\qquad
		D_k(\hat O)=\operatorname{Tr}(\hat O\hat r_k)
		\;,\\
		w_k(\hat O,\hat X)&=D_k(\hat O)E_k(\hat X)
		\;.
	\end{aligned}
	\label{eq:input_readout_overlaps}\alsolabel{eq:generic_modal_weight}
\end{equation}
For a simple eigenvalue, a reciprocal rescaling of the matched eigenoperators changes the excitation and detection factors but leaves their product and the spectral projector invariant [Eq.~\eqref{eq:biorthogonal_gauge_transformation}]. For an initial state $\hat\rho_{\rm in}$, the evolving density matrix and the deviation of an observable from its stationary mean are
\begin{equation}
	\begin{aligned}
		\hat\rho(t)&=\hat\rho_{\rm ss}
		+\sum_{k>0}e^{\lambda_k t}E_k(\hat\rho_{\rm in})\hat r_k
		\;,\\
		\delta\langle\hat O\rangle_t
		&\equiv\operatorname{Tr}[\hat O(\hat\rho(t)-\hat\rho_{\rm ss})]
		=\sum_{k>0}e^{\lambda_k t}w_k(\hat O,\hat\rho_{\rm in})
		\;.
	\end{aligned}
	\label{eq:one_shot_amplitude}\alsolabel{eq:density_modal_expansion}\alsolabel{eq:switching_coherent_amplitude}
\end{equation}
The excitation factor sets the initial amplitude of a mode, the exponential its time dependence, and the detection factor its contribution to the chosen observable. A preparation with $E_k=0$ eliminates the mode from the evolving state, whereas an observable with $D_k=0$ is insensitive to the mode even when it is present. A faster mode with a larger modal weight can dominate the measured relaxation at finite times. Figure~\ref{fig:scheme1} illustrates these distinctions for a Kerr switching mode. The same factorization holds for stationary correlations and linear response (Appendix~\ref{app:correlations_response}). Because $E_k$ is linear in the state, its average over the conditional states of any unravelling of the master equation into quantum trajectories equals $E_k(\hat\rho(t))=E_k(\hat\rho_{\rm in})e^{\lambda_kt}$. Trajectory-resolved weights, used to select Liouvillian eigenvalues in Ref.~\cite{Ferrari2025}, can depend on the unravelling through their moduli and fluctuations.

\subsection{Phase-space excitation maps and reconstruction}
\label{subsec:phase_space_representation}

Coherent states provide a continuous family of preparations parametrized by $\alpha(X,P)=(X+iP)/\sqrt{2}$, analogous to the globally rotated spin preparations of Ref.~\cite{Kochsiek2022}. With $\mathcal W$ denoting the Wigner transform (Appendix~\ref{app:phase_space_conventions}), we define
\begin{equation}
	\begin{aligned}
		W_{\rm ss}(X,P)&=\mathcal W(\hat\rho_{\rm ss};X,P)
		\;,\\
		W_k^R(X,P)&=\mathcal W(\hat r_k;X,P)
		\;,\\
		\mathcal E_k(X,P)&\equiv E_k(|\alpha\rangle\langle\alpha|)
		=\langle\alpha|\hat\ell_k^\dagger|\alpha\rangle
		\;.
	\end{aligned}
	\label{eq:left_susceptibility_map}\alsolabel{eq:right_mode_wigner_symbol}\alsolabel{eq:switching_left_right_maps}
\end{equation}
The excitation map $\mathcal E_k(X,P)=\pi Q_{\hat\ell_k^\dagger}(\alpha(X,P))$ is proportional to the Husimi symbol of the left eigenoperator, whereas $W_k^R(X,P)$ describes the contribution of the right eigenoperator to the evolving Wigner function~\cite{Cahill1969,Hillery1984}. Equivalently, $\mathcal E_k$ is the Weyl symbol of $\hat\ell_k^\dagger$ averaged over the Wigner function of $|\alpha\rangle$, a Gaussian of variance $1/2$ per quadrature. This average smooths the symbol of the left eigenoperator on the scale of a coherent state. The two maps represent different operators through different phase-space transforms, so their widths and amplitudes are not directly comparable. For a simple real eigenvalue, the matched eigenoperators can be chosen Hermitian, making the excitation map real (Appendix~\ref{app:phase_space_conventions}).

The zeros of $\mathcal E_k$ identify coherent preparations that do not excite mode $k$. A coherent preparation that does not excite a unique real slowest mode can relax faster than an initially closer state that excites it [Fig.~\ref{fig:scheme1}(f)]~\cite{Lu2017,Carollo2021}. The same zeros can also limit the reconstruction of the slow generator from a single preparation (Sec.~\ref{sec:v11identification}).

The excitation map can be reconstructed from transient measurements. Fitting the relaxation of an observable $\hat O$ for different coherent preparations with a common set of resolved poles $\{\lambda_k\}$ gives modal amplitudes $A_k^{(O)}(X,P)=D_k(\hat O)\mathcal E_k(X,P)$. Provided that $D_k(\hat O)\neq0$ and the amplitudes can be resolved, the excitation map is determined up to a nonzero complex factor per mode. Its zeros and relative amplitudes are then accessible without calculating the left eigenoperator. Additional observables determine relative detection factors for the same mode (Appendix~\ref{app:rc_measured}).

We call a slow sector separated when $\Gamma_{\rm rest}/\Gamma_{\rm sector}\geq10$, where $\Gamma_{\rm rest}$ is the smallest decay rate outside the retained stationary-plus-slow sector and $\Gamma_{\rm sector}$ the largest retained one. For nearly degenerate slow eigenvalues, we use the propagator restricted to the common invariant subspace of these modes~\cite{Kato1995,Albert2014,Minganti2019,Pan2026} (Appendix~\ref{app:resolved_sectors}).

\section{Symmetry-resolved relaxation in Kerr resonators}
\label{sec:driven_kerr}

Driven Kerr resonators exhibit a variety of metastable configurations controlled by the external drives~\cite{Drummond1980,Bartolo2016,Minganti2018,Minganti2023,Gravina2023,Beaulieu2025,Roberts2020}. We first consider linear and parametric driving, for which the slow decay rates and stationary populations determine the effective transition rates. Section~\ref{subsec:biased_kerr_kinetic_rerouting} examines what remains of this correspondence when parity is broken.

\subsection{Model and preparation protocol}
\label{subsec:kerr_model_protocol}

We consider a single bosonic mode $\hat a$ with Kerr nonlinearity $U$, one-photon drive $F$ at frequency $\omega_d$, and two-photon drive $G$ at $2\omega_d$, as realized in superconducting microwave circuits~\cite{Puri2020,Grimm2020}. In the frame rotating at $\omega_d$ and with $\hbar=1$, the Hamiltonian is
\begin{equation}
	\hat H=-(\Delta-U)\hat a^\dagger\hat a
	+\frac{U}{2}\hat a^{\dagger 2}\hat a^2
	+F\hat a^\dagger+F^*\hat a
	+\frac{G}{2}\hat a^{\dagger 2}+\frac{G^*}{2}\hat a^2
	\;,
	\label{eq:kerr_hamiltonian}
\end{equation}
and the density matrix evolves according to
\begin{equation}
	\partial_t\hat\rho=\mathcal L\hat\rho
	=-i[\hat H,\hat\rho]+\kappa\mathcal D(\hat a)\hat\rho
	\;.
	\label{eq:kerr_master_equation}
\end{equation}
Here $\kappa$ is the one-photon loss rate and $\mathcal D(\hat L)\hat\rho=\hat L\hat\rho\hat L^\dagger-\frac12\{\hat L^\dagger\hat L,\hat\rho\}$ is the Lindblad dissipator. The detuning $\Delta=\omega_d-\omega_0+U$, with $\omega_0$ the resonator frequency, includes the Kerr shift (Appendix~\ref{app:kerr_shift}). Coherent displacements prepare a chosen amplitude and phase, and quadrature and photon-number measurements distinguish the metastable configurations [Fig.~\ref{fig:scheme1}(a)].

We introduce the photon-number scale $\aleph$ through $U=\widetilde U/\aleph$ and $F=\sqrt{\aleph}\,\widetilde F$, at fixed $\Delta$, $G$, $\kappa$, $\widetilde U$, and $\widetilde F$, so that coherent amplitudes scale as $\alpha=\sqrt\aleph\,\widetilde\alpha$~\cite{Casteels2017a,Seibold2020,Nowoczyn2026a,Nowoczyn2026}. The drift of the Wigner function is then the scaled semiclassical flow
\begin{equation}
	i\partial_t\widetilde\alpha
	=\Big(-\Delta-\frac{i\kappa}{2}\Big)\widetilde\alpha
	+\widetilde U|\widetilde\alpha|^2\widetilde\alpha
	+\widetilde F+G\widetilde\alpha^*
	\label{eq:semiclassical_flow}
\end{equation}
at every $\aleph$, while diffusion and the third-order Kerr term scale as $\aleph^{-1}$ and $\aleph^{-2}$ (Appendix~\ref{app:kerr_shift}). Increasing $\aleph$ thus reduces quantum fluctuations at a fixed flow~\cite{Nowoczyn2026a}, and $\aleph\to\infty$ is the semiclassical limit.

For $F=0$, the Liouvillian commutes with the parity superoperator $\mathcal Z_2(\hat\rho)=\hat\Pi\hat\rho\hat\Pi^\dagger$, where $\hat\Pi=\exp(i\pi\hat n)$. The Hamiltonian and dissipator are invariant under $\hat\Pi\hat a\hat\Pi^\dagger=-\hat a$, although individual photon-loss events change photon-number parity. A nonzero one-photon drive breaks this weak symmetry (Appendix~\ref{app:spectral_conventions}).

\subsection{Linearly driven Kerr: one switching mode}
\label{subsec:one_photon_switching_basin}

For $F\neq0$ and $G=0$, the semiclassical bistable regime contains a low-amplitude (dim) and a high-amplitude (bright) response. At finite $\aleph$, a separated real Liouvillian mode with decay rate $\Gamma_{\rm sw}=-\operatorname{Re}(\lambda_{\rm sw})$ describes slow switching between these metastable configurations~\cite{Drummond1980,Vogel1989,Kheruntsyan1999,Minganti2018,Roberts2020,Carde2026}. When switching is much slower than the relaxation within each configuration, its right eigenoperator is, to leading order, proportional to the difference between the two metastable states.

Let $p_{\rm hi}(\hat\rho)$ and $p_{\rm lo}(\hat\rho)=1-p_{\rm hi}(\hat\rho)$ denote the coordinates of the bright and dim representatives in $\hat\rho_{\rm ss}+\operatorname{span}(\hat r_{\rm sw})$. Choosing the extreme eigenvalues $\ell_{\rm hi}$ and $\ell_{\rm lo}$ of the Hermitian left switching eigenoperator to define these coordinates, we obtain
\begin{equation*}
	\mathcal E_{\rm sw}(X,P)
	=(\ell_{\rm hi}-\ell_{\rm lo})
	[p_{\rm hi}(X,P)-\pi_{\rm hi}]
	\;.
\end{equation*}
Here $p_{\rm hi}(X,P)=p_{\rm hi}(|\alpha\rangle\langle\alpha|)$ and $\pi_{\rm hi}=p_{\rm hi}(\hat\rho_{\rm ss})$. The excitation zero identifies preparations that already carry the stationary bright-state coordinate; it coincides with the equal-weight line $p_{\rm hi}=1/2$ only when $\pi_{\rm hi}=1/2$. Across the detuning-induced crossover from bright to dim stationary occupation, $\pi_{\rm hi}$ falls from nearly one to nearly zero. The zero contour then moves from the bright towards the dim attractor, while the right-mode Wigner symbol changes little (Appendix~\ref{app:detuning_scans}, Fig.~\ref{fig:kerr_bistable_switching_tomography}).

In Fig.~\ref{fig:scheme1}(f), the preparation on the zero contour starts farther from stationarity but relaxes on the scale $\Gamma_2^{-1}$ and overtakes the preparation that excites the switching mode and relaxes on the scale $\Gamma_1^{-1}$. This comparison exhibits the strong Mpemba effect~\cite{Lu2017,Klich2019,Carollo2021,Zhang2025}. The excitation map is not itself a probability, and its zeros need not coincide with an equal-committor surface or a deterministic basin separatrix~\cite{Macieszczak2021,Brown2024,Weinan2010}.

For an admissible two-state Markov reduction, the decay rate is the sum of the two transition rates, $\Gamma_{\rm sw}=k_{{\rm hi}\leftarrow{\rm lo}}+k_{{\rm lo}\leftarrow{\rm hi}}$, while stationarity gives $\pi_{\rm hi}=k_{{\rm hi}\leftarrow{\rm lo}}/\Gamma_{\rm sw}$. Hence
\begin{equation*}
	k_{{\rm hi}\leftarrow{\rm lo}}=\pi_{\rm hi}\Gamma_{\rm sw}
	\;,\qquad
	k_{{\rm lo}\leftarrow{\rm hi}}=(1-\pi_{\rm hi})\Gamma_{\rm sw}
	\;.
\end{equation*}

\subsection{Parametrically driven Kerr: odd and even slow modes}
\label{subsec:two_photon_parity_bright_vacuum}

For $F=0$ and $G\neq0$, two symmetry-related finite-amplitude configurations can coexist with a central low-amplitude configuration. Their stability follows from the driven, damped semiclassical flow. Because this flow is not a gradient flow, quasienergy barriers alone do not determine their transition rates.

Parity covariance allows the Liouvillian eigenoperators to be chosen even or odd under $\mathcal Z_2$. This classification differs from the photon-number parity of cat states. In the investigated parameter window, two real modes with decay rates $\Gamma_o=-\operatorname{Re}(\lambda_o)$ and $\Gamma_e=-\operatorname{Re}(\lambda_e)$ separate from the faster spectrum~\cite{Gravina2023,Minganti2023,Beaulieu2025,Carde2026,Ruiz2023,Frattini2024}. We label them by parity, not by eigenvalue ordering.

The odd mode describes the population imbalance between the outer lobes. This imbalance decays through both inter-lobe switching, including quantum activation~\cite{Thompson2026}, and capture into the center, so $\Gamma_o$ is not the inter-lobe switching rate alone. The latter exhibits a stepwise suppression in two-photon-driven oscillators due to the interplay of tunnelling, dephasing, and decay~\cite{Frattini2024,Su2025}. The even mode describes the population contrast between the center and the two lobes together. When escape from the center is negligible, its decay rate predominantly measures center capture~\cite{Gravina2023}. As the stationary occupation moves with detuning from the outer lobes to the center, the odd mode changes little, while the largest even-mode excitation amplitudes move from the center to the outer lobes (Appendix~\ref{app:detuning_scans}, Figs.~\ref{fig:cat_detuning_scan} and~\ref{fig:cat_transition_tomography}). These maps can be probed with operations available in Kerr-parametric-oscillator and Kerr-cat experiments~\cite{Grimm2020,Suzuki2023}.

\paragraph{Three-state reduction.}

We label the central configuration $c$ and the outer lobes $+$ and $-$ according to $\langle P\rangle_+>0$ and $\langle P\rangle_-<0$. Parity exchanges the lobes and leaves the center invariant, allowing three independent rates: escape from the center into either lobe at rate $a/2$, capture from either lobe at rate $b$, and inter-lobe switching at rate $s$. In the ordered basis $(c,+,-)$, the column-conserving generator is
\begin{equation}
	\mathbf Q_{\mathbb Z_2}
	=
	\begin{pmatrix}
		-a & b & b\\
		a/2 & -(b+s) & s\\
		a/2 & s & -(b+s)
	\end{pmatrix}
	\;.
	\label{eq:cat_symmetric_three_state_generator}
\end{equation}
Its odd and even eigenvectors are $\mathbf v_o\propto(0,1,-1)^{\mathsf T}$ and $\mathbf v_e\propto(1,-1/2,-1/2)^{\mathsf T}$ (Appendix~\ref{app:symmetric_three_state}), with eigenvalues
\begin{equation}
	\lambda_o=-(b+2s)
	\;,\qquad
	\lambda_e=-(a+b)
	\;.
	\label{eq:cat_symmetric_nonstationary_rates}
\end{equation}
Inter-lobe switching reverses the odd population imbalance but leaves the even contrast unchanged, explaining the factor of two in $\lambda_o$ and the absence of $s$ from $\lambda_e$. The stationary weights are
\begin{equation}
	\pi_c=\frac{b}{a+b}
	\;,\qquad
	\pi_+=\pi_-=\frac{a}{2(a+b)}
	\;.
	\label{eq:cat_symmetric_stationary_weights}
\end{equation}

To construct the reduction from the Liouvillian, we select three candidate representatives $\hat\sigma_i$, $i\in\{c,+,-\}$, from the extremal geometry of the projected physical-state set within $\hat\rho_{\rm ss}+\operatorname{span}(\hat r_o,\hat r_e)$ (Appendix~\ref{app:slow_coordinate_construction}). Writing $\hat\sigma_i=\sum_\mu C_{\mu i}\hat r_\mu$, with $\mu\in\{{\rm ss},o,e\}$, $\hat r_{\rm ss}=\hat\rho_{\rm ss}$, and $C_{{\rm ss},i}=1$, defines the dual operators $\hat\eta_j^\dagger=\sum_\mu(C^{-1})_{j\mu}\hat\ell_\mu^\dagger$ and coordinates $p_j(\hat\rho)=\operatorname{Tr}(\hat\eta_j^\dagger\hat\rho)$~\cite{Macieszczak2021}. These coordinates satisfy $\sum_jp_j(\hat\rho)=1$ for normalized states. We write $\pi_j=p_j(\hat\rho_{\rm ss})$ and $p_j(X,P)=p_j(|\alpha\rangle\langle\alpha|)$.

The generator in this basis is obtained directly from the Liouvillian, without fitting transition rates~\cite{Macieszczak2021,Rose2022},
\begin{equation}
	\mathbf Q=C^{-1}\Lambda_{\rm slow}C
	\;,\qquad
	\Lambda_{\rm slow}=\operatorname{diag}(0,\lambda_o,\lambda_e)
	\;,
	\label{eq:cat_projected_generator}
\end{equation}
with $Q_{ji}=\operatorname{Tr}[\hat\eta_j^\dagger\mathcal L(\hat\sigma_i)]$. Equivalently,
\begin{equation*}
	Q_{ji}
	=\sum_{\mu=o,e}\lambda_\mu
	\operatorname{Tr}(\hat\eta_j^\dagger\hat r_\mu)
	\operatorname{Tr}(\hat\ell_\mu^\dagger\hat\sigma_i)
	\;.
\end{equation*}
As in Eq.~\eqref{eq:one_shot_amplitude}, the generator entries are sums of products of left and right overlaps. Fixing the eigenvalues does not fix these products. Because the dual operators lie in the invariant left slow subspace, $\mathcal L^\dagger\hat\eta_j^\dagger=\sum_iQ_{ji}\hat\eta_i^\dagger$ for real $\mathbf Q$, so $p(t)=e^{t\mathbf Q}p(0)$ holds exactly for every initial state, including coherent preparations far from any attractor. The faster modes must have decayed only for reconstructing $\hat\rho(t)$ from its coordinates and for estimating the coordinates from readout means.

The coordinates are not necessarily probabilities, nor are the representatives necessarily positive density matrices. A stochastic interpretation requires nonnegative coordinates for the states under consideration and a real, column-conserving generator with nonnegative off-diagonal entries (Appendix~\ref{app:slow_coordinate_checks}). Even then, its transition rates are not necessarily first-passage rates of individual quantum trajectories. When stochastic admissibility fails, $\mathbf Q$ still describes the affine slow dynamics. Three semiclassical attractors alone therefore do not guarantee an admissible three-state reduction at finite $\aleph$: spectral separation, positivity of the coordinates, and stochasticity of $\mathbf Q$ must be tested separately.

For a parity-covariant representative construction, $\mathbf Q$ has the form of Eq.~\eqref{eq:cat_symmetric_three_state_generator}, and Eqs.~\eqref{eq:cat_symmetric_nonstationary_rates} and~\eqref{eq:cat_symmetric_stationary_weights} give
\begin{equation}
	a=(1-\pi_c)\Gamma_e
	\;,\qquad
	b=\pi_c\Gamma_e
	\;,\qquad
	s=\frac{\Gamma_o-\pi_c\Gamma_e}{2}
	\;.
	\label{eq:cat_symmetric_inversion}
\end{equation}
For physical stationary weights and positive decay rates, $\Gamma_o\geq\pi_c\Gamma_e$ is necessary and sufficient for nonnegative rates. Parity fixes the directions of both nonstationary coordinate eigenvectors, so the two decay rates and the stationary central weight determine the generator uniquely. Breaking parity removes this constraint (Sec.~\ref{sec:rc}). The even excitation map is $\mathcal E_e(X,P)\propto p_c(X,P)-\pi_c$, so, as in the two-state case, its zero marks preparations with the stationary central coordinate and imposes no further constraint on the rates.

For a nearly absorbing center, $\pi_c\simeq1$, Eq.~\eqref{eq:cat_symmetric_inversion} gives $a\ll b$ and $b\simeq\Gamma_e$. The approximation $s\simeq(\Gamma_o-\Gamma_e)/2$ additionally requires $(1-\pi_c)\Gamma_e\ll|\Gamma_o-\Gamma_e|$. In the strictly absorbing limit,
\begin{equation*}
	\frac{s}{b+s}
	=\frac{\Gamma_o-\Gamma_e}{\Gamma_o+\Gamma_e}
	\;,
\end{equation*}
so nearly degenerate odd and even decay rates imply dominant center capture.

At the zero-bias operating point of Sec.~\ref{subsec:biased_kerr_kinetic_rerouting}, $(\Delta,G)/\widetilde U=(5,0.4)$, the ratio $b/s$ increases from $122$ at $\aleph=5$ to $241$ at $\aleph=7$ and $804$ at $\aleph=10$. At $\aleph=5$, the relative pole splitting is $(\Gamma_o-\Gamma_e)/(\Gamma_o+\Gamma_e)=0.0081$, and both lobes predominantly decay into the center.

\section{Competing transitions under symmetry breaking}
\label{subsec:biased_kerr_kinetic_rerouting}

We break parity with a signed real one-photon drive $F$, keeping $G$ real and working around the zero-bias operating point
\begin{equation}
	\Delta/\widetilde U=5
	\;,\qquad
	G/\widetilde U=0.4
	\;.
	\label{eq:biased_operating_point}
\end{equation}
At zero bias, the two outer lobes predominantly decay into the nearly absorbing center (Sec.~\ref{subsec:two_photon_parity_bright_vacuum}). The bias changes the competition between center capture and inter-lobe switching. These transitions describe relaxation between metastable configurations, not logical errors of an idling Kerr-cat qubit. We retain the labels $c$, $+$, and $-$ for the center and the lobes with positive and negative $\langle P\rangle$, respectively, and track the two slow modes from their zero-bias parity eigenoperators using biorthogonal overlaps.

Although parity is broken at fixed nonzero $F$, bias reversal gives $\mathcal L(-F)=\mathcal Z_2\mathcal L(F)\mathcal Z_2^{-1}$. The spectra at opposite biases are identical. Isolated, nondegenerate eigenvalues continued analytically from $F=0$ are even in $F$, so their first-order shifts vanish. For a parity-covariant coordinate construction that varies smoothly with bias, by contrast, the difference between the two inter-lobe branching probabilities is odd in $F$: the branching asymmetry can change at first order in the bias, whereas isolated slow decay rates change only at second order or higher.

\begin{figure}[t]
	\centering
	\includegraphics[width=0.94\linewidth]{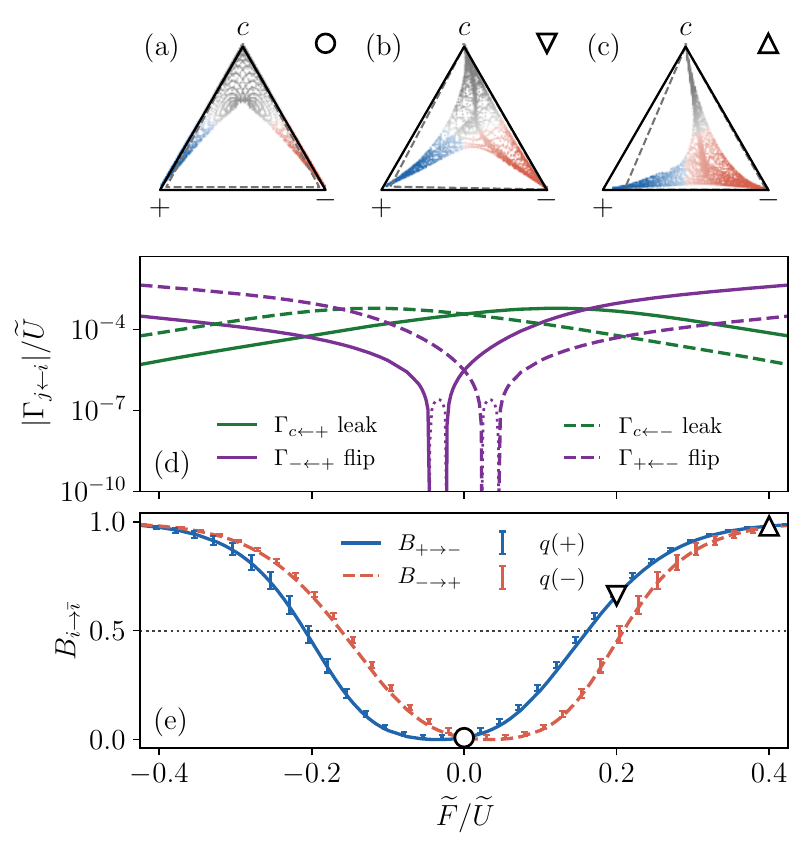}
	\caption{\textit{Competing metastable transitions under a one-photon bias.} (a--c) Slow sector in representative coordinates $(p_c,p_+,p_-)$ at $\widetilde F/\widetilde U=0$, $0.2$, and $0.4$. Corners denote the representatives, solid outlines the coordinates accessible to physical states, dashed outlines positive reconstructed slow states, and dots coherent preparations. The positivity defect $\varepsilon_\sigma$ of these left-accessible representatives (the magnitude of their most negative eigenvalue) increases from $0.020$ to $0.025$ and $0.088$, while $\Gamma_{\rm rest}/\Gamma_{\rm sector}$ decreases from $133$ to $35$ and $12$. (d) Moduli of the prepared-basis entries for center capture $\Gamma_{c\leftarrow\pm}$ (green) and inter-lobe switching $\Gamma_{\mp\leftarrow\pm}$ (purple), from $+$ (solid) and $-$ (dashed). All resolved off-diagonal entries are positive except $\Gamma_{+\leftarrow-}$ for $0.023<\widetilde F/\widetilde U<0.045$ and, by bias reversal, $\Gamma_{-\leftarrow+}$ for $-0.045<\widetilde F/\widetilde U<-0.023$ (dotted), where the entry is negative with modulus below $9\times10^{-4}$ of the competing capture entry; the cusps are the interpolated zero crossings. (e) Projected branching fractions $B_{\pm\to\mp}$ (lines; probabilities only where the reduction is Markov) and full quantum first-detection probabilities $q(\pm)$ (bars) for detection near the opposite lobe before detection near the center. The bars cover $0.1\leq\gamma/\widetilde U\leq1$ and scaled detector radii $r=0.5,0.75$. Data are not interpolated. The separation criterion $\Gamma_{\rm rest}/\Gamma_{\rm sector}\geq10$ holds for $|\widetilde F|/\widetilde U\leq0.425$. Parameters: $\Delta/\widetilde U=5$, $G/\widetilde U=0.4$, $\kappa/\widetilde U=0.1$, $\aleph=5$, and $N=60$.}
	\label{fig:biased_kerr_directional_branching}
\end{figure}

\paragraph{Prepared coordinates and transition rates.}

The geometric representatives in Fig.~\ref{fig:biased_kerr_directional_branching}(a--c) are not positive density matrices. We therefore introduce an experimentally accessible coordinate frame using coherent preparations at the three mean-field attractors. Each preparation is evolved to a plateau time $t_0$, beyond the initial fast transient, and its slow coordinates form one column of $P_0$. For affinely independent anchors, the generator in the prepared basis is $\mathbf Q'=P_0^{-1}\mathbf QP_0$. The evolved preparations are physical states, but their slow projections need not be positive, and the transformation does not guarantee a Markov generator.

The quadrature and photon-number readout pairs of the three anchors at $t_0$ define an affine calibration when the vectors $(\langle \hat X_\phi\rangle,\langle\hat n\rangle,1)^{\mathsf T}$ are linearly independent. Residual contributions of the fast modes limit the accuracy of this calibration. In the ordered basis $(c,+,-)$, we write
\begin{equation}
	Q'_{ji}=\Gamma_{j\leftarrow i}\quad(j\neq i)
	\;,\qquad
	Q'_{ii}=-\sum_{k\neq i}\Gamma_{k\leftarrow i}
	\;.
	\label{eq:biased_three_state_generator_pm}
\end{equation}
The off-diagonal entries are transition rates only where the stochastic admissibility conditions of Appendix~\ref{app:slow_coordinate_checks} hold. Center escape remains below numerical resolution throughout the scan. Appendix~\ref{app:biased_asymmetries} compares the prepared and representative bases and documents their behavior under bias reversal.

Where both competing exit rates are positive, we characterize their competition by the log odds $\mathcal R_\pm$; for a nonzero total exit rate, we define the branching fractions $B_{i\to j}$,
\begin{equation}
	\mathcal R_\pm=\ln\!\left(
	\frac{\Gamma_{\mp\leftarrow\pm}}{\Gamma_{c\leftarrow\pm}}
	\right)
	\;,\qquad
	B_{i\to j}=
	\frac{\Gamma_{j\leftarrow i}}{\sum_{k\neq i}\Gamma_{k\leftarrow i}}
	\;.
	\label{eq:biased_routing_log_odds}\alsolabel{eq:biased_outer_branching_fractions_pm}
\end{equation}
For an admissible Markov generator and nonzero total exit rate, $B_{i\to j}$ is the probability that the next transition from $i$ leads to $j$; we then call it a branching probability, and otherwise a projected branching fraction. When both competing rates are positive, $B_{\pm\to\mp}=(1+e^{-\mathcal R_\pm})^{-1}$. Positive (negative) log odds indicate a preferred exit to the opposite lobe (to the center). An asymmetry between the switching rates alone does not determine the preferred exits, because the capture rates may also differ.

\paragraph{Bias-dependent preferred exits.}

At zero bias, both lobes prefer center capture [Fig.~\ref{fig:biased_kerr_directional_branching}(d,e)]. A small positive bias strongly suppresses switching from $-$ to $+$. In the prepared basis, $\Gamma_{+\leftarrow-}$ changes sign at $\widetilde F/\widetilde U\simeq0.023$ and $0.045$ and reaches $-2.5\times10^{-7}\,\widetilde U$ in between, at most $9\times10^{-4}$ of the competing capture rate. In the representative basis it remains positive, so in this window the sign of the entry depends on the basis (Appendix~\ref{app:biased_asymmetries}). The independent quantum detection probability $q(-)$ has a minimum in the same interval. This supports a directional suppression of switching but does not establish a rate zero. Related suppression has been reported in biased two-state parametric oscillators~\cite{Boness2025}.

At larger bias, the two log odds cross zero near $\widetilde F/\widetilde U=0.16$ and $0.21$; across the tested reductions and detection settings, the crossover intervals are $[0.158,0.163]$ and $[0.197,0.213]$. At the reference bias $\widetilde F/\widetilde U=0.2$, the prepared-basis branching probabilities are $B_{+\to-}=0.661$ and $B_{-\to+}=0.450$. Every procedure in Table~\ref{tab:routing_spread} gives the $+$ lobe a preference for switching. The $-$ lobe is close to its crossover. Alternative representative constructions give $B_{-\to+}$ up to $0.471$, and the quantum detection probabilities reach $0.520$ for the detector settings of Fig.~\ref{fig:biased_kerr_directional_branching}(e), so its preferred exit at this bias is not resolved (Table~\ref{tab:routing_spread}). Beyond both crossings, switching is preferred from either lobe, and reversing the bias exchanges the roles of the lobes. At the three biases of Table~\ref{tab:routing_spread}, the detection probabilities for these settings lie within $0.07$ of the projected branching probabilities. They reproduce the crossovers but depend on the instrument, so they are not detector-independent first-passage probabilities.

The preparation coordinates describe how an initial state populates the metastable configurations, whereas $\mathbf Q'$ describes how these configurations then exchange population. Preparation regions defined by the largest representative coordinate, and the mean-field basins, rearrange near the first crossover (Appendix~\ref{app:biased_asymmetries}, Fig.~\ref{fig:biased_kerr_membership_rerouting}). Neither determines the quantum branching probabilities. The second crossover has no corresponding mean-field basin reconnection, and the classical reconnections at $\widetilde F/\widetilde U=0.249$ and $0.331$ leave the preferred exits unchanged.

\paragraph{Excitation of the two slow modes.}

\begin{figure}[t]
	\centering
	\includegraphics[width=0.80\linewidth]{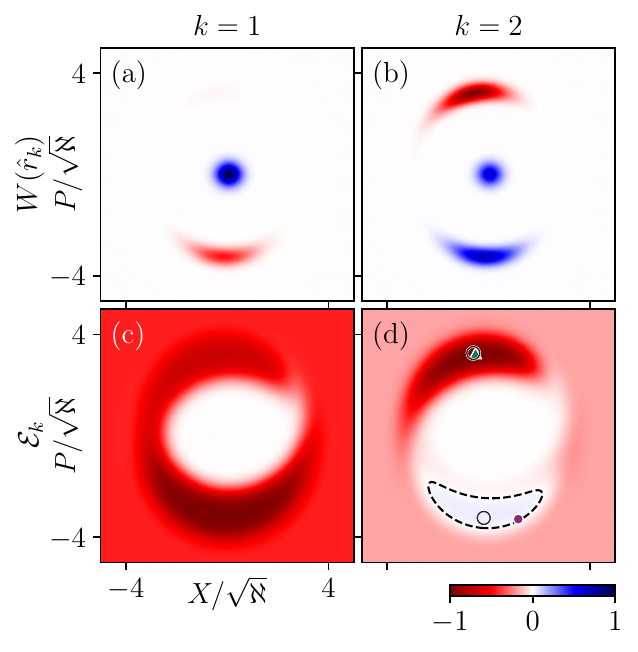}
	\caption{\textit{Slow modes of the biased Kerr resonator at $\widetilde F/\widetilde U=0.2$.} (a,b) Right-mode Wigner symbols $W_k^R(X,P)$, with $\lambda_1/\widetilde U=-7.5\times10^{-5}$ and $\lambda_2/\widetilde U=-1.4\times10^{-3}$. (c,d) Corresponding coherent-state excitation maps $\mathcal E_k(X,P)$. Each panel is normalized to its maximum modulus, with signs fixed by $\operatorname{Tr}(\hat n\hat r_1)<0$ and $\operatorname{Tr}(\hat P\hat r_2)<0$. The dashed contour in (d) marks $\mathcal E_2=0$; inside, $\mathcal E_2$ is positive and reaches at most $0.038$ of its maximum modulus. Open circles denote the lobe anchors, the purple dot a preparation on the calculated excitation-zero contour, and the green triangle a preparation exciting both modes. Both selected preparations are checked independently at $N=80$. Spectral separation $\Gamma_{\rm rest}/\Gamma_{\rm sector}=35$, $N=230$; other parameters as in Fig.~\ref{fig:biased_kerr_directional_branching}.}
	\label{fig:biased_kerr_lr_tomography}
\end{figure}

Figure~\ref{fig:biased_kerr_lr_tomography} shows the two slow modes at $\widetilde F/\widetilde U=0.2$. The right eigenoperator $\hat r_1$ predominantly exchanges population between the center and the $-$ lobe, while $\hat r_2$ describes redistribution between the $+$ lobe and the other configurations. Both outer lobes excite the slower mode, whereas appreciable excitation of the faster mode occurs only in the $+$ region. For the coherent inputs with the largest coordinate margin in each region, $\mathcal E_1(-)/\mathcal E_1(+)=1.43$ and $\mathcal E_2(-)/\mathcal E_2(+)=-0.038$; for the mean-field anchors used in Sec.~\ref{sec:v11identification}, the same ratios are $1.43$ and $-0.037$. The $-$ anchor lies inside the $\mathcal E_2=0$ contour of Fig.~\ref{fig:biased_kerr_lr_tomography}(d): a preparation of the $-$ lobe excites almost exclusively the slower mode. Section~\ref{sec:v11identification} shows that this weak excitation of the faster mode makes the full generator expensive to reconstruct from the $-$ lobe.

\section{Spectral constraints on metastable transitions}
\label{sec:rc}

\subsection{Constraints on competing transition probabilities}

\begin{figure}[!tp]
	\centering
	\includegraphics[width=\linewidth]{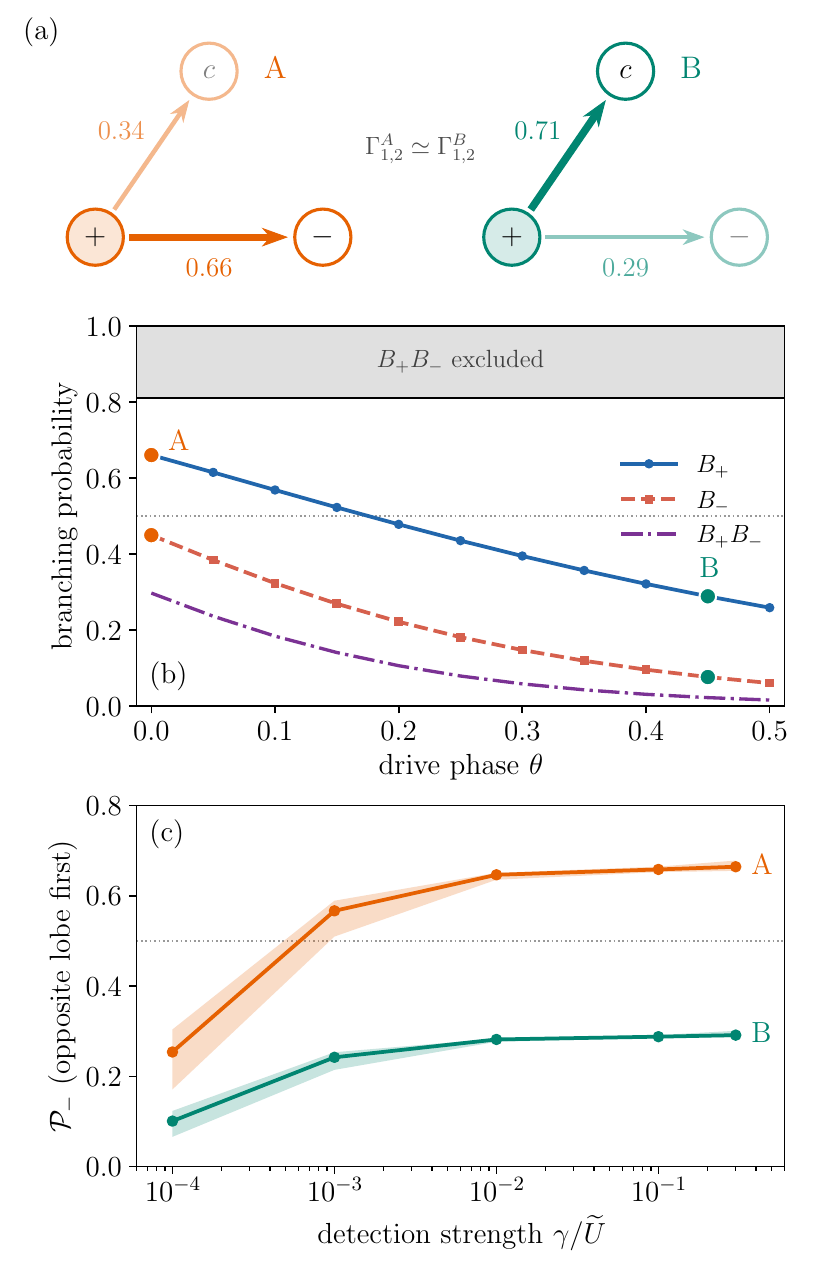}
	\caption{\textit{Opposite preferred exits at numerically matched slow poles.} (a) Three-state networks at $A$ ($\theta=0$) and $B$ ($\theta=0.45$), arranged as the coordinate triangles of Figs.~\ref{fig:biased_kerr_directional_branching} and~\ref{fig:v11identifiability}. The arrows from the $+$ lobe carry the projected branching probabilities, with widths growing with their values; the preferred exit is drawn in full color. Both slow decay rates agree at $A$ and $B$, and at both points the center is nearly absorbing, $\pi_c\simeq1$. (b) Inter-lobe branching probabilities $B_\pm=B_{\pm\to\mp}$ and their product along the drive-phase continuation at $|\widetilde F|/\widetilde U=0.2$, with $\Delta$ and $G$ adjusted to keep both slow decay rates fixed. The absorbing-center bound of Eq.~\eqref{eq:rc_absorbing_bound} is therefore the constant $0.811$ set by $\Gamma_2/\Gamma_1=19.15$ (black line). In the absorbing-center limit, the shaded region is excluded for the product, not for $B_+$ or $B_-$ individually. (c) Probability $\mathcal P_-$ that the two-channel instrument detects the opposite lobe before the center, after preparing the $+$ lobe, for equal channel strengths $\gamma_c=\gamma_-=\gamma$. Lines: scaled detector radius $r=0.75$ and the reference input; bands: range over $r=0.50$, $0.75$, $1.00$ and five input displacements. Each calculation uses its own cutoff, $N=80$ (generators), $110$ (continuation) and $140$ (instrument); between these cutoffs the detection probabilities change by less than $5\times10^{-7}$ (Appendix~\ref{app:v11endpoints}).}
	\label{fig:rc_kernel}
\end{figure}

In the parity-symmetric three-state model, symmetry fixes the nonstationary coordinate eigenvectors, allowing the transition rates to be reconstructed from the decay rates and stationary populations. Without parity, prescribing two distinct real decay rates fixes only the trace and determinant of the generator restricted to the zero-mass plane. The remaining freedom changes how population leaving an outer lobe is distributed between the opposite lobe and the center. Related ambiguities arise in aggregated Markov and competing-risk models~\cite{Siekmann2026,Lindqvist2023}. Neither the ambiguity nor the bound derived below relies on quantum coherence, and even the full Liouvillian spectrum need not remove the ambiguity. Consider the classical rate model with jump operators $\sqrt{Q_{ji}}\,|j\rangle\langle i|$ and, in the basis $(c,+,-)$, the generator $\Gamma[(0,0,0)^{\mathsf T},(4-4x,-4,4x)^{\mathsf T},(1,0,-1)^{\mathsf T}]$. For every $0\leq x\leq1$ it has the same stationary state, the same total exit rate of each lobe, and the same complete spectrum, $\{0,-\Gamma,-4\Gamma\}$ for the populations and $\{-\Gamma/2,-2\Gamma,-5\Gamma/2\}$, each twofold, for the coherences, while $B_{+\to-}=x$. The quantum part of the problem is how such a slow sector emerges from a nonlinear resonator and how coherent preparations and quantum readouts give access to its rates.

Consider a stochastic three-state reduction with an absorbing center and two positive decay rates $\Gamma_1<\Gamma_2$. Let $u=-Q_{+,+}$ and $v=-Q_{-,-}$ be the total lobe exit rates, and $B_+=B_{+\to-}$ and $B_-=B_{-\to+}$ the respective inter-lobe branching probabilities. The restriction of the generator to the two lobes is
\begin{equation*}
	\mathbf T=
	\begin{pmatrix}
		-u & B_-v\\
		B_+u & -v
	\end{pmatrix}
	\;.
\end{equation*}
Its eigenvalues are $-\Gamma_1$ and $-\Gamma_2$. With $S=\Gamma_1+\Gamma_2$ and $D=\Gamma_1\Gamma_2$, the trace and determinant give
\begin{equation}
	u+v=S
	\;,\qquad
	uv(1-B_+B_-)=D
	\;.
\end{equation}
The discriminant of $\mathbf T$ then splits the squared relative pole separation into two nonnegative contributions,
\begin{equation*}
	\left(\frac{\Gamma_2-\Gamma_1}{\Gamma_2+\Gamma_1}\right)^2
	=B_+B_-+(1-B_+B_-)\left(\frac{u-v}{u+v}\right)^2
	\;:
\end{equation*}
round trips between the lobes and unequal lobe lifetimes. A small separation therefore limits repeated switching between the lobes, while a large one need not imply it. Dropping the second term gives
\begin{equation}
	B_+B_-\leq
	\left(
	\frac{\Gamma_2-\Gamma_1}{\Gamma_2+\Gamma_1}
	\right)^2
	\;.
	\label{eq:rc_absorbing_bound}
\end{equation}
This bound is sharp: for every pair $0\leq B_\pm\leq1$ satisfying Eq.~\eqref{eq:rc_absorbing_bound}, positive exit rates $u,v$ exist with sum $S$ and product $D/(1-B_+B_-)$. Splitting each exit rate according to its branching probability constructs the corresponding Markov generator. Equality holds if and only if $u=v$.

The product $B_+B_-$ is the probability of completing a round trip between the lobes before center capture, starting from either lobe. The number of completed round trips $N_{\rm rt}$ satisfies $\Pr(N_{\rm rt}\geq m)=(B_+B_-)^m$, giving
\begin{equation*}
	\langle N_{\rm rt}\rangle
	=
	\frac{B_+B_-}{1-B_+B_-}
	\leq
	\frac{(\Gamma_2-\Gamma_1)^2}{4\Gamma_1\Gamma_2}
	\;.
\end{equation*}
Nearly degenerate decay rates suppress repeated switching before capture, without determining either branching probability individually. Both lobes can preferentially switch to each other ($B_\pm>1/2$) for some admissible generator if and only if $\Gamma_2>3\Gamma_1$. At zero bias, parity gives $u=v$, saturating the bound in the absorbing-center limit; the nearly absorbing zero-bias Kerr resonator approaches this equality with $B_+=B_-\simeq0.0081$.

With escape from the center and prescribed stationary populations $\boldsymbol\pi=(\pi_c,\pi_+,\pi_-)^{\mathsf T}$, the bound generalizes to the necessary condition $B_+B_-\leq1-4\pi_c\Gamma_1\Gamma_2/(\Gamma_1+\Gamma_2)^2$, and a discriminant condition with nonnegative escape rates is necessary and sufficient (Appendix~\ref{app:rc_kernel_checks}). At a finite-return Kerr point, the branching product $0.381$ exceeds the absorbing-center value $0.139$ but lies inside this feasible region [Fig.~\ref{fig:v12bias_checks}(c)]. The matched Kerr resonators considered below have $\pi_c\simeq1$, with center escape below numerical resolution, so Eq.~\eqref{eq:rc_absorbing_bound} is the relevant constraint.

\subsection{Opposite preferred transitions at matched slow poles}

To realize different preferred exits at fixed decay rates, we rotate the one-photon drive, $\widetilde F=|\widetilde F|e^{i\theta}$, keeping $|\widetilde F|/\widetilde U=0.2$, $\aleph=5$, and $\kappa/\widetilde U=0.1$. At each phase, we adjust $\Delta$ and $G$ to keep the slow decay rates of the operating point $A$, $(\theta,\Delta/\widetilde U,G/\widetilde U)=(0,5,0.4)$, namely $\Gamma_1/\widetilde U=7.479\times10^{-5}$ and $\Gamma_2/\widetilde U=1.432\times10^{-3}$. Along this continuation, the preferred exit from the $+$ lobe reverses near $\theta=0.18$ [Fig.~\ref{fig:rc_kernel}(a,b)]. At the operating point $B$, $(\theta,\Delta/\widetilde U,G/\widetilde U)\simeq(0.45,8.920,0.410)$, the projected branching probability is $B_+=0.289$, compared with $0.661$ at $A$. The total exit rate of the $+$ lobe changes only from $1.40$ to $1.43$ in units of $10^{-3}\widetilde U$, while its flip rate falls from $0.92$ to $0.41$ and its capture rate rises from $0.47$ to $1.02$. The preferred exit thus reverses at a nearly unchanged mean residence time. In the prepared-basis generator, the $-$ lobe prefers the center throughout the continuation, with $B_-$ falling from $0.450$ at $A$ to $0.076$ at $B$. The product $B_+B_-$ falls from $0.30$ to $0.02$, far below the bound $0.811$ set by $\Gamma_2/\Gamma_1=19.15$ [Fig.~\ref{fig:rc_kernel}(b)]. By the discriminant identity above, unequal lobe lifetimes account for the rest of the squared relative splitting, $0.51$ at $A$ and $0.79$ at $B$: the mean residence time of the $-$ lobe is $13$ times that of the $+$ lobe at $A$ and $19$ times at $B$. The opposite preferred exits persist under anchor perturbations and alternative constructions using positive representatives (Appendix~\ref{app:v11closure}).

At $N=140$, the computed decay rates at $A$ and $B$ differ by less than $5\times10^{-9}$ relative, and refitting at larger cutoff reduces the difference below $10^{-10}$. Both mismatches lie below the eigenpair residuals of Table~\ref{tab:endpoint}; they quantify agreement between the computed poles, not their absolute accuracy. At both points the evolved central anchor is numerically indistinguishable from its stationary state, so $\pi_c\simeq1$ and the stationary coordinates agree as well. The matching thus concerns the slow eigenvalues and the stationary populations in the respective calibrated frames, not the stationary density operators. These, the faster poles, and the lobe preparations differ between $A$ and $B$; the plateau mean photon number of the $+$ lobe is $23.3$ at $A$ and $35.7$ at $B$. The negligible center escape is a property of this coordinate construction, not an independently resolved rate (Appendix~\ref{app:biased_asymmetries}). Individual transition rates depend on the coordinate frame. The reversal of the preferred exit, however, persists across the representative constructions tested here, and the detection instrument below resolves it independently.

To test the reversal without a three-state reduction, we introduce a quantum instrument with two absorbing detection channels associated with disjoint phase-space disks $\Omega_j$ centered on the central and negative-lobe attractors. Their positive detection effects are
\begin{equation}
	\hat E_j=
	\int_{\Omega_j}\frac{\mathrm d^2\alpha}{\pi}\,
	\hat P_N|\alpha\rangle\langle\alpha|\hat P_N
	\;,\qquad j=c,-
	\;,
	\label{eq:v11effects}
\end{equation}
where $\hat P_N$ projects onto the retained Fock levels and $\hat E_c+\hat E_-\leq\hat I$. For jumps into the absorbing channels with strengths $\gamma_c$ and $\gamma_-$, the unnormalized state conditioned on no detection evolves under
\begin{equation}
	\mathcal K\hat\rho=
	\mathcal L\hat\rho
	-\frac12\{\gamma_c\hat E_c+\gamma_-\hat E_-,\hat\rho\}
	\;.
\end{equation}
The first-detection probability from the local positive-lobe preparation $\hat\rho_+$ is $\mathcal P_j=\operatorname{Tr}(\hat F_j\hat\rho_+)$, with $\hat F_j=\gamma_j\int_0^\infty e^{t\mathcal K^\dagger}\hat E_j\,\mathrm dt$, or equivalently $\mathcal K^\dagger \hat F_j=-\gamma_j\hat E_j$. This calculation uses the full density operator~\cite{Chia2017}. The protocol requires monitored loss into distinguishable absorbing channels (Appendix~\ref{app:v11endpoints}), and the conditional evolution includes measurement backaction before the first detection, so the probabilities depend on the instrument.

At scaled detector radius $r=0.75$ and equal strengths $\gamma_c=\gamma_-=0.01\widetilde U$, the opposite-lobe detection probabilities are $0.647$ at $A$ and $0.282$ at $B$ [Fig.~\ref{fig:rc_kernel}(c)], reproducing the reversal. The reversal persists for the three radii and five tested inputs at detection strengths from $10^{-3}$ to $0.3$ in units of $\widetilde U$. At $10^{-4}\widetilde U$, both resonators instead favor center detection. For equal strengths $\gamma$ and a relaxing unmonitored Liouvillian, the weak-detection limit is
\begin{equation*}
	\lim_{\gamma\to0}\mathcal P_-
	=
	\frac{\operatorname{Tr}(\hat E_-\hat\rho_{\rm ss})}
	{\operatorname{Tr}[(\hat E_c+\hat E_-)\hat\rho_{\rm ss}]}
	\;,
\end{equation*}
provided that the total stationary detector weight is nonzero: slow detection probes the stationary state, not the initial exit preference. At half the damping, a separately matched pair retains the reversal at a weaker detector strength (Appendix~\ref{app:v11endpoints}).

\section{Operational reconstruction of slow-sector dynamics}
\label{sec:v11identification}

The resonators $A$ and $B$ have numerically matched slow decay rates and stationary coordinates, yet their $+$ lobes prefer different exits. The matched slow poles do not distinguish these exit preferences, but calibrated transient responses do. We now determine which transition rates can be inferred from one preparation and two measured means, and how many shots this requires. The object inferred is the calibrated slow generator; we do not attempt to identify a microscopic Lindbladian from general time records.

\subsection{Identifying the preferred exit from a single delay}

\begin{figure}[!tp]
	\centering
	\includegraphics[width=\linewidth]{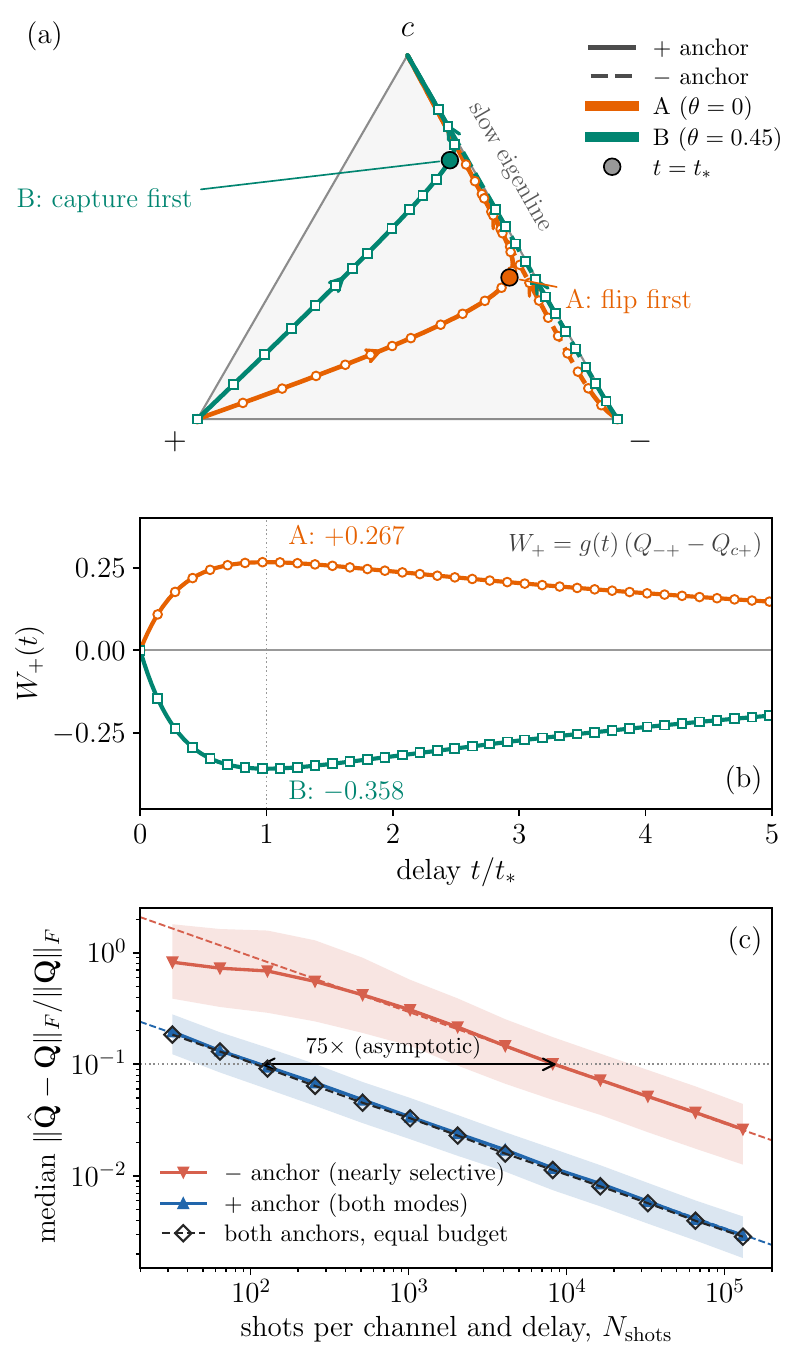}
	\caption{\textit{Reconstruction of slow-sector dynamics from transient measurements.} (a) Slow-sector trajectories at the matched points $A$ (orange) and $B$ (teal), each in its own calibrated coordinates $(p_c,p_+,p_-)$, center at the top. Solid (dashed): $+$ ($-$) anchor; curves: Eq.~\eqref{eq:rc_kernel}; open symbols: full Lindblad evolution ($N=80$); filled circles: $t=t_*$; arrowheads: direction of time. (b) Stationary-corrected contrast $W_+(t)$ of the $+$ anchor [Eq.~\eqref{eq:rc_witness}]; curves: $g(t)\Delta k_+$; symbols as in (a). (c) Median relative Frobenius error of the generator reconstructed at $A$ from $N_{\rm shots}$ outcomes per readout channel at $t_0$ and $t_*$, with exact poles and calibration ($4000$ trials; bands: interquartile range), for the $-$ and $+$ anchors (cyclic estimator) and for both anchors with $N_{\rm shots}/2$ outcomes each. All estimators impose the known poles. Dashed lines: $c/\sqrt{N_{\rm shots}}$ with $c=9.3$ and $1.08$. The arrow marks the asymptotic shot ratio $(9.3/1.08)^2\simeq75$ at $10\%$ median error ($64$ on the sampled grid).}
	\label{fig:v11identifiability}
\end{figure}

We first infer the preferred exit without reconstructing the full generator. Let $Q$ denote the generator in the prepared coordinates $(c,+,-)$, with $\mathbf 1^{\mathsf T}Q=0$ and $Q\pi=0$, where $\pi$ is the normalized stationary coordinate vector. With $\Pi=\pi\mathbf 1^{\mathsf T}$, interpolation at the stationary eigenvalue and the two distinct real slow eigenvalues gives~\cite{Putzer1966}
\begin{align}
	e^{tQ}
	&=\Pi+h(t)(I-\Pi)+g(t)Q
	\;,
	\label{eq:rc_kernel}\\
	g(t)
	&=\frac{e^{-\Gamma_1t}-e^{-\Gamma_2t}}{\Gamma_2-\Gamma_1}
	\;,\qquad
	h(t)=e^{-\Gamma_1t}+\Gamma_1g(t)
	\;.
\end{align}
For two generators with identical slow decay rates and the same calibrated initial coordinates $p_0$, the transient difference is
\begin{equation*}
	p_A(t)-p_B(t)
	=[1-h(t)](\pi_A-\pi_B)
	+g(t)(Q_A-Q_B)p_0
	\;.
\end{equation*}
When the stationary coordinates coincide, all coordinate differences follow the same kernel $g(t)$, with amplitudes $(Q_A-Q_B)p_0$ (Appendix~\ref{app:rc_kernel_checks}, Fig.~\ref{fig:v16_kernel_collapse}). With $P_1$ the rank-one spectral projector of $Q$ onto its slowest mode, $Q=-\Gamma_2(I-\Pi)+(\Gamma_2-\Gamma_1)P_1$: once $\pi$ and the poles are fixed, $Q$ is determined by $P_1$, and $p_A(t)-p_B(t)=(e^{-\Gamma_1t}-e^{-\Gamma_2t})(P_{1,A}-P_{1,B})p_0$. The coefficient $g(t)$ reaches its maximum at
\begin{equation}
	t_*=\frac{\ln(\Gamma_2/\Gamma_1)}{\Gamma_2-\Gamma_1}
	\;,\qquad
	g(t_*)=\frac{e^{-\Gamma_1t_*}}{\Gamma_2}
	\;,
	\label{eq:rc_kernel_peak}
\end{equation}
which is $\kappa t_*\simeq217$ for the matched resonators. The delay $t_*$ maximizes $g(t)$; the statistically optimal delay also depends on how measurement noise and calibration errors vary with the delay (Appendix~\ref{app:slow_coordinate_checks}).

Figure~\ref{fig:v11identifiability}(a) shows the transients in the calibrated slow coordinates. The faster mode decays first and leaves the population of the prepared $+$ lobe near the slow eigenline, which nearly coincides with the edge between $c$ and $-$. The population accumulated in the $-$ lobe reflects the competition between the two exits: near $t_*$, the opposite-lobe coordinate reaches $p_-\simeq0.55$ at $A$ but only $0.25$ at $B$. This transient population is not a branching probability, since population that has reached the $-$ lobe can leak into the center or switch back. The population then relaxes along the eigenline into the center.

To measure this, we record the rotated quadrature $\langle \hat X_\phi\rangle$ and photon number $\langle\hat n\rangle$ on separate ensembles at a delay $t$ after the plateau time $t_0$. The three anchor readouts at $t_0$, supplemented by normalization, form the columns of the calibration matrix $M$ (Sec.~\ref{subsec:biased_kerr_kinetic_rerouting}). The coordinates are $p(t)=M^{-1}(\langle \hat X_\phi\rangle,\langle\hat n\rangle,1)^{\mathsf T}$, and the same transformation of the stationary readouts gives $\pi$. This requires known poles, an invertible calibration, and negligible contributions from the faster modes. For the positive-lobe anchor, $p(0)=e_+$, and Eq.~\eqref{eq:rc_kernel} determines the entire generator column from one calibrated delay,
\begin{equation*}
	Qe_+ = \frac{p(t)-\pi-h(t)(e_+-\pi)}{g(t)}
	\;,
\end{equation*}
including both exit entries $Q_{j,+}=\{p_j(t)-[1-h(t)]\pi_j\}/g(t)$ for $j=-,c$. The preferred exit follows from their difference,
\begin{align}
	W_+(t)
	&=p_-(t)-p_c(t)-(\pi_--\pi_c)[1-h(t)]
	\notag\\
	&=g(t)\Delta k_+
	\;,\qquad
	\Delta k_+=Q_{-,+}-Q_{c,+}
	\;.
	\label{eq:rc_witness}
\end{align}
For an admissible Markov generator, $\Delta k_+=-Q_{+,+}(2B_+-1)$, so the sign of $\Delta k_+$ identifies the preferred exit; outside the Markov region it remains a difference of affine generator entries [Fig.~\ref{fig:v12bias_checks}(b)]. Since $g(t)>0$ for every $t>0$, only the sign of $W_+$ is needed, at any delay. The poles enter $W_+$ only through the stationary correction $(\pi_--\pi_c)[1-h(t)]$, so a pole error changes the sign only if the error it induces in this correction is comparable with $|W_+|$. The column $Qe_+$, by contrast, depends on the poles through both $h(t)$ in the numerator and $g(t)$ in the denominator (Appendix~\ref{app:rc_measured}). The rate contrast $\Delta k_+$ differs from the dimensionless log odds $\mathcal R_+$ of Eq.~\eqref{eq:biased_routing_log_odds}.

The projected generators predict $W_+(t_*)=+0.267$ at $A$ and $-0.358$ at $B$, and the calibrated Lindblad transients follow these predictions [Fig.~\ref{fig:v11identifiability}(b)]. With calibration, drift, and pole uncertainties and $80$ delayed shots per readout channel, the direct estimate recovers the sign of $\Delta k_+$ in $99.8\%$ and $100\%$ of $5000$ simulated trials at $A$ and $B$ (Appendix~\ref{app:v11inference}).

\subsection{Reconstructing the full generator from one preparation}

For a general preparation, let $v_0=p(0)-\pi$ denote its initial deviation from stationarity. Given $v_0$, Eq.~\eqref{eq:rc_kernel} determines $q=Qv_0$ from the coordinates at one positive delay. On the plane $\mathbf 1^{\mathsf T}v=0$, the Cayley--Hamilton theorem gives $Q^2+SQ+DI=0$, with $S=\Gamma_1+\Gamma_2$ and $D=\Gamma_1\Gamma_2$. Hence $Qq=-Sq-Dv_0$. Together with $Q\pi=0$, this yields
\begin{align}
	q&=\frac{p(t)-\pi-h(t)v_0}{g(t)}
	\;,
	\notag\\
	Q&=[\,0,\ q,\ -Sq-Dv_0\,]
	[\,\pi,\ v_0,\ q\,]^{-1}
	\;,
	\label{eq:v11cyclic}
\end{align}
provided that the three columns of the inverted matrix are linearly independent. One calibrated delay determines $Qv_0$, and the known characteristic polynomial then gives $Q(Qv_0)$; this is the cyclic-vector construction for a two-dimensional dynamical subspace~\cite{Kalman1963,HoKalman1966}. Alternatively, measurements from a second preparation with an independent deviation from $\pi$ determine the action of $Q$ along a second direction directly (Appendix~\ref{app:v11inference}).

The excitation maps of Fig.~\ref{fig:biased_kerr_lr_tomography} determine whether the cyclic basis is invertible. Let $r_k$ denote the coordinate representation of $\hat r_k$ and write $v_0=\mathcal E_1r_1+\mathcal E_2r_2$, with excitation amplitudes evaluated at the plateau origin. Relative to the initial coherent preparation, these amplitudes acquire factors $e^{-\Gamma_k t_0}$. Using $Qr_k=-\Gamma_kr_k$, we obtain
\begin{equation}
	\det[\pi,v_0,Qv_0]
	=(\Gamma_1-\Gamma_2)
	\mathcal E_1\mathcal E_2
	\det[\pi,r_1,r_2]
	\;.
	\label{eq:v11det}
\end{equation}
The determinant is invariant under reciprocal eigenoperator rescalings. Consider the unconstrained real affine family of generators with prescribed stationary coordinates, conserved total weight, and two known distinct real slow poles. Within this family, one preparation and one positive delay identify the generator if and only if the preparation excites both modes. With $m\geq2$ distinct real slow poles and full calibrated coordinates, one preparation and $m-1$ positive delays suffice exactly when the preparation excites all $m$ modes, $\prod_{k=1}^m\mathcal E_k\neq0$ (Appendix~\ref{app:v11inference}). Additional structural constraints can restrict this family further. In system-realization approaches to quantum system identification, an initial state aligned with an eigenvector of the dynamical generator was already noted to leave the output insensitive to the unknown parameters~\cite{ZhangSarovar2015}, and identifiability conditions for autonomous and controlled open quantum systems have been formulated through linear and bilinear system identification~\cite{Parvaiz2025}. Equations~\eqref{eq:v11det} and~\eqref{eq:v12eigenline_task} make this precise for a calibrated two-mode metastable sector with known poles: they give the exact criterion for one preparation and one delay and specify which rate combinations remain identifiable when it fails.

At an excitation zero with $v_0\neq0$, the slow trajectory stays on a single eigenline, so additional delays cannot probe the action of $Q$ along an independent direction. A preparation exciting both modes probes two independent directions because the relative modal amplitudes change during relaxation. At $A$, the two lobe anchors approximate these two cases [Fig.~\ref{fig:v11identifiability}(a)]: the $+$ trajectory bends from the fast to the slow eigendirection, whereas the $-$ trajectory runs approximately along the slow eigenline from the start. In unit-normalized coordinate eigenvectors, the anchor deviations at the plateau origin are $v_0=0.98\,r_1+1.22\,r_2$ for $+$ and $v_0=1.40\,r_1-0.045\,r_2$ for $-$. The $-$ anchor is thus nearly, but not exactly, selective, and its determinant in Eq.~\eqref{eq:v11det} is $5.3\%$ of that of the $+$ anchor. Reconstructing the full generator from this anchor is therefore possible in principle but poorly conditioned (Sec.~\ref{subsec:measurement_cost}).

An exactly single-mode preparation still determines part of the kinetics. For a nonzero single-mode deviation $v_0$, the compatible generators differ by $\delta Q=\zeta v_0f^{\mathsf T}$, where $\zeta$ is a free real parameter and $f$ is a fixed nonzero covector satisfying $f^{\mathsf T}\pi=f^{\mathsf T}v_0=0$ (Appendix~\ref{app:v11inference}). The ambiguity in a scalar quantity $d^{\mathsf T}Qw$ is $\zeta(d^{\mathsf T}v_0)(f^{\mathsf T}w)$. Such a quantity is identifiable within the unconstrained affine family if and only if
\begin{equation}
	d^{\mathsf T}v_0=0
	\quad\text{or}\quad
	w\in\operatorname{span}\{\pi,v_0\}
	\;.
	\label{eq:v12eigenline_task}
\end{equation}
For a lobe anchor that excites only one mode, $e_\pm=\pi+v_0$ belongs to the measured subspace. Such a preparation fixes its own column, and with it its exit contrast and preferred exit, but leaves the action of the generator along the other slow direction undetermined. For the nearly selective $-$ anchor this column, including $\Delta k_-=Q_{+,-}-Q_{c,-}$, is directly measured by the single-delay formula above, whereas the remaining action of $Q$ is only weakly constrained.

A small excitation of the faster mode does not imply a small exit rate. Write the deviation of a lobe anchor as $e_i-\pi=z_1+z_2$, with $z_k$ its component along $r_k$. Since $Qz_k=-\Gamma_kz_k$, its column is $Qe_i=-\Gamma_1(e_i-\pi)-(\Gamma_2-\Gamma_1)z_2$, so that $Q_{ji}=\Gamma_1\pi_j-(\Gamma_2-\Gamma_1)(z_2)_j$ for $j\neq i$. With $\pi_+\simeq0$, the whole switching rate of the $-$ anchor at $A$ comes from its faster-mode component: the admixture $0.045\,r_2$, multiplied by $\Gamma_2-\Gamma_1=18.2\,\Gamma_1$, gives $Q_{+,-}=4.9\times10^{-5}\,\widetilde U$ and $B_-=0.45$. Nearly single-mode relaxation from a lobe is therefore compatible with a sizable inter-lobe branching probability, and the small component that carries this rate is the one that makes the full generator expensive to reconstruct from that lobe.

\subsection{Measurement cost of selective excitation}
\label{subsec:measurement_cost}

Near an excitation zero, cyclic reconstruction becomes sensitive to measurement noise. We quantify its geometric conditioning by the smallest singular value $\sigma_{\min}$ of $[v_0,Qv_0/S]$ in an orthonormal basis of the zero-mass plane; dividing by $S$ gives both columns the same units. This diagnostic is independent of eigenoperator normalization but depends on the coordinate frame. Near the slow eigenline it measures the distance from it: to first order in $\mathcal E_2$, $\sigma_{\min}=(\Gamma_2-\Gamma_1)(S^2+\Gamma_1^2)^{-1/2}\,d_{\rm line}$, where $d_{\rm line}=|\mathcal E_2|\,\|r_2\|\,|\sin(\vartheta)|$ is the Euclidean distance of $v_0$ from $\operatorname{span}(r_1)$, or equivalently of $p(0)$ from the eigenline $\pi+\operatorname{span}(r_1)$, and $\vartheta$ is the angle between $r_1$ and $r_2$. At $A$, the $+$ anchor has $\sigma_{\min}=0.618$ and the $-$ anchor $0.040$, which the formula reproduces from $d_{\rm line}=0.045$; at $B$, the $-$ anchor is even more selective, with $\sigma_{\min}=0.0047$.

To estimate the number of measurement repetitions, we simulate the protocol at $A$ [Fig.~\ref{fig:v11identifiability}(c)]. We sample $N_{\rm shots}$ quadrature and $N_{\rm shots}$ photon-number outcomes at $t_0$ and at $t_*$ from the exact finite-Fock outcome distributions after full Lindblad evolution and reconstruct $Q$ with Eq.~\eqref{eq:v11cyclic}, keeping the poles and the calibration at their exact values. For the $+$ anchor, the median relative Frobenius error follows $1.08/\sqrt{N_{\rm shots}}$ from $N_{\rm shots}=32$ on. For the $-$ anchor it approaches $9.3/\sqrt{N_{\rm shots}}$ only above $N_{\rm shots}\simeq300$; below, the errors are of order one and the inversion abstains in up to $4\%$ of the trials. For the same median error, the asymptotic scaling requires $(9.3/1.08)^2\simeq75$ times more shots per channel and delay from the $-$ lobe, $8.7\times10^3$ instead of $1.2\times10^2$ for $10\%$; on the sampled grid the simulated medians give a factor $64$. Sampling the initial coordinates barely changes this factor: with the anchor coordinates at $t_0$ supplied exactly, $v_0=e_\pm-\pi$, the prefactors become $8.6$ and $1.00$ and the ratio $74$ (Appendix~\ref{app:v11inference}). The squared singular-value ratio $(0.618/0.040)^2\simeq240$ overestimates the cost, because $\sigma_{\min}$ is a geometric diagnostic: the statistical error also depends on the measurement covariance and on its orientation relative to the sensitive directions of the estimator.

With $\pi$, the poles, and the calibration fixed, the generator can still vary along two shears $\delta Q=\delta a\,r_1l_2^{\mathsf T}+\delta b\,r_2l_1^{\mathsf T}$, where $l_k$ are the covectors dual to $r_k$ on the zero-mass plane. These directions preserve the poles to first order, the product $\delta a\,\delta b$ shifts them at second order, and the transient responds as
\begin{equation}
	\delta p(t)=g(t)\,\delta Q\,v_0
	=g(t)\left(\delta a\,\mathcal E_2r_1+\delta b\,\mathcal E_1r_2\right)
	\;.
	\label{eq:isospectral_sensitivity}
\end{equation}
The coupling $\delta a$ is visible only through the excitation of the faster mode. In a protocol that estimates the readout means $y=M_2p$, with single-shot covariance $\Sigma$ and readout directions $m_k=M_2r_k$, the asymptotic Fisher information per shot on $\delta a$, with $\delta b$ unknown, is the Schur complement
\begin{equation}
	\mathcal I_{a|b}=g(t)^2\,\mathcal E_2^2\left(m_1^{\mathsf T}\Sigma^{-1}m_1-\frac{(m_1^{\mathsf T}\Sigma^{-1}m_2)^2}{m_2^{\mathsf T}\Sigma^{-1}m_2}\right)
	\;,
	\label{eq:schur_information}
\end{equation}
which separates the temporal, preparation, and readout factors. When the temporal and readout factors remain finite and nonzero, the number of shots needed for a given precision on $\delta a$ therefore grows as $|\mathcal E_2|^{-2}$, whereas the contribution of the mode to the relaxation is linear in $\mathcal E_2$. Near a regular zero contour, $|\mathcal E_2|\simeq|\nabla\mathcal E_2|\,d_\perp$, where $d_\perp$ is the normal distance from the contour, and the shot number grows as $d_\perp^{-2}$. This distance can be estimated without the left eigenoperator, as $|A_2^{(O)}|/|\nabla A_2^{(O)}|$ from the measured amplitudes of Sec.~\ref{subsec:phase_space_representation}. On the pilot grid, $\mathcal I_{a|b}$ scales as $|\mathcal E_2|^{2.03}$ and as $d_\perp^{1.94}$ for the $49$ preparations within $0.1\sqrt\aleph$ of the zero contour, and the smallest eigenvalue of the full Fisher matrix as $|\mathcal E_2|^{2.07}$ on $219$ preparations near it (Appendix~\ref{app:v11inference}). With known initial coordinates, the cyclic estimator inverts a square map from two measured means to the two generator parameters. Its asymptotic covariance attains the Cram\'er--Rao bound for these means, which gives median-error prefactors $8.61$ and $1.00$. The resulting shot-cost ratio of $74$ thus agrees with the asymptotic information bound for the measured means at $t_*$. The full outcome distributions can carry additional information in their higher moments, which this protocol does not use.

Splitting the same total budget between the two anchors and fitting the two-preparation estimate under the same known-pole constraint gives the prefactor $1.03$, close to the $1.08$ of the $+$ anchor alone. The $-$ anchor adds little because the $+$ lobe already excites both modes. Without the pole constraint, the two-preparation formula of Appendix~\ref{app:v11inference} gives $1.53$, so comparisons between preparation strategies must use equally constrained estimators. These costs characterize the calibrated protocol at $A$ and are not lower bounds on the cost of learning the generator.

Transients also identify the preparations that excite both modes. Excitation maps reconstructed from simulated quadrature and photon-number transients of $21\times49$ coherent preparations locate both the faster-mode zero near the $-$ lobe and the well-conditioned region near the $+$ lobe. At both operating points, the calibrated slow evolution of independent test preparations agrees with the full density matrices to trace distance below $2\times10^{-8}$ (Appendices~\ref{app:v11closure} and~\ref{app:v11inference}, Fig.~\ref{fig:pilot_map}).

In units of the Kerr coefficient, the reference point $A$ at $\aleph=5$ has $\kappa/U=0.5$, $G/U=2$, $|F|/U\simeq2.24$, and $\Delta_{\rm bare}/U=24$, where $\Delta_{\rm bare}=\Delta-U=\omega_d-\omega_0$ is the bare detuning (Appendix~\ref{app:kerr_shift}). Point $B$ has $\Delta_{\rm bare}/U\simeq43.6$ and $G/U\simeq2.05$, so $A$ and $B$ correspond to different drive and detuning settings of the same device. The calibration is taken at $\kappa t_0=25$ and the delayed readout after a further $\kappa t_*\simeq217$. The shot costs above exclude the measurements needed to determine the poles, the stationary coordinates, and the excitation maps.

\section{Conclusion and outlook}
\label{sec:discussion_outlook}

We have developed an operational spectral theory of Liouvillians that connects mode excitation, state deformation, and readout through matched left and right eigenoperators. Coherent-state excitation maps identify preparations that suppress individual modes and can be reconstructed from resolved transients with nonzero readout overlap. For the switching mode, their zeros mark the stationary population balance, not necessarily a basin boundary. Within a metastable slow sector, the matched eigenoperators also define operational coordinates and, for stochastic reductions, a transition generator.

In driven Kerr resonators, the slow decay rates and stationary populations fix the two-state and the parity-symmetric three-state Markov generators. When parity is broken, the rates are in general underdetermined, but for an absorbing center the slow decay rates sharply bound the inter-lobe round-trip probability and the mean number of round trips before capture. Rotating the drive phase, with compensating changes of the detuning and parametric drive, lowers the inter-lobe branching probability of a lobe from $0.661$ to $0.289$, reversing its preferred exit at numerically matched slow decay rates and nearly unchanged mean residence time. A quantum detection model reproduces the reversal without a three-state reduction.

Given two distinct real slow poles, stationary coordinates, and an invertible readout calibration, two observable means at one delay determine both exit rates of a prepared lobe without resolving rare switching events, and the sign of a stationary-corrected contrast gives its preferred exit. Within the unconstrained affine generator family, the same data identify the full generator if and only if the preparation excites both slow modes. Near a regular zero contour of the faster mode's excitation map, the shot cost at fixed precision grows as the inverse squared distance, if the other information factors stay finite and nonzero. At the reference point, the nearly selective negative lobe asymptotically requires about $75$ times as many shots as the positive lobe at equal median error. Preparation geometry thus serves two complementary aims: suppressing the slowest mode can accelerate relaxation, as in the strong Mpemba effect, whereas identifying the full generator from one preparation requires exciting both slow modes.

The reversal is demonstrated at $\kappa/K=1$, with $K=U/2$, above the damping ratios of the cited Kerr-cat devices~\cite{Grimm2020,Frattini2024,CrosCarrilloDeAlbornoz2024}. At half this damping ratio, resolving it depends on the detector timescale (Appendix~\ref{app:v11endpoints}). Appendix~\ref{app:v11inference} extends the reconstruction criteria to larger manifolds and complex poles. Numerical applications of these extensions and a systematic study at weaker damping remain open. The broader question of what Liouvillian eigenoperators reveal beyond their eigenvalues extends to the reconstruction of classical flow topology, to metastability in collective systems, and to dissipative quantum chaos.

\begin{acknowledgments}
I gratefully acknowledge funding from the Deutsche Forschungsgemeinschaft (DFG, German Research Foundation) through the Heisenberg Programme (Project No.~449653034) and SFB 1432 (Project No.~425217212).
I thank D.~Z.~Haxell and N.~Beato for helpful feedback on the manuscript and C.~Nowoczyn for fruitful discussions.
\end{acknowledgments}

\FloatBarrier
\appendix

\section{Spectral conventions and resolved sectors}
\label{app:spectral_framework}

\subsection{Eigensystem, mode gauge, and parity}
\label{app:spectral_conventions}

We work in a finite-dimensional Liouville space with the unique stationary state of Eq.~\eqref{eq:steady_state}; irreducibility conditions provide sufficient criteria~\cite{Spohn1977,Evans1977,Frigerio1978}. For the finite-cutoff systems considered here, all nonstationary eigenvalues have strictly negative real parts. Hermiticity preservation pairs nonreal eigenvalues with their complex conjugates. Biorthonormality, Eq.~\eqref{eq:right_left_eigenoperators}, leaves the reciprocal freedom
\begin{equation}
	\hat r_k\rightarrow a_k\hat r_k
	\;,\qquad
	\hat\ell_k\rightarrow(a_k^{-1})^*\hat\ell_k
	\;,\qquad
	a_k\in\mathbb C\setminus\{0\}
	\;,
	\label{eq:biorthogonal_gauge_transformation}
\end{equation}
which we call the mode gauge. It is not a physical symmetry: eigenspaces, spectral projectors, and the left-right products used in the main text are invariant under it.

At the finite $\aleph$ and $\kappa>0$ used here the stationary state is unique, and the stationary states, retained spectral data, and phase-space maps are stable under enlargement of the Fock cutoff. For $F=0$ the Liouvillian commutes with $\mathcal Z_2$; this weak symmetry supports the two-lobe Kerr-cat slow sector~\cite{Buca2012,Puri2020,Grimm2020,Gravina2023,Minganti2023}, and a one-photon drive breaks it and mixes the symmetry-adapted modes.

\subsection{Exact Wigner equation and Kerr-shifted detuning}
\label{app:kerr_shift}

In the scaled quadratures $\widetilde X=X/\sqrt\aleph$ and $\widetilde P=P/\sqrt\aleph$, with $\widetilde\alpha=(\widetilde X+i\widetilde P)/\sqrt2$, the Wigner function of the untruncated oscillator obeys
\begin{equation}
	\partial_tW=-\nabla\cdot(\mathbf fW)+\frac{\kappa}{4\aleph}\nabla^2W
	-\frac{\widetilde U}{8\aleph^2}\big(\widetilde X\partial_{\widetilde P}-\widetilde P\partial_{\widetilde X}\big)\nabla^2W
	\;,
	\label{eq:wigner_equation}
\end{equation}
where $\nabla=(\partial_{\widetilde X},\partial_{\widetilde P})$ and $\mathbf f$ is the real form of the flow of Eq.~\eqref{eq:semiclassical_flow} (see below). Equation~\eqref{eq:wigner_equation} is exact because the Moyal expansion of the quartic Hamiltonian terminates at third order. At finite $\aleph$, the mean of $\widetilde\alpha$ evolves with the Wigner average of the nonlinear drift, not with the drift at the mean; Eq.~\eqref{eq:semiclassical_flow} governs it as $\aleph\to\infty$ for states that remain localized in the scaled phase space.

The detuning in Eq.~\eqref{eq:kerr_hamiltonian}, $\Delta=\omega_d-\omega_0+U$, includes a Kerr shift, as in Ref.~\cite{Seibold2026}. The bare detuning of the drive from the resonator frequency is $\Delta_{\rm bare}=\omega_d-\omega_0=\Delta-U$. With $\hat n=\hat a^\dagger\hat a$, the undriven Hamiltonian is $-\Delta\hat n+U\hat n(\hat n+1)/2=-\Delta_{\rm bare}\hat n+(U/2)\hat a^{\dagger2}\hat a^2$. The Weyl symbols $(\hat a^\dagger\hat a)_W=|\alpha|^2-1/2$ and $(\hat a^{\dagger2}\hat a^2)_W=|\alpha|^4-2|\alpha|^2+1/2$ give it the Weyl symbol $-\Delta|\alpha|^2+(U/2)|\alpha|^4$ up to a constant. The term $U\hat a^\dagger\hat a$ thus cancels the quadratic Weyl correction $-U|\alpha|^2$ of the normal-ordered Kerr term. Together with the damping from $\kappa\mathcal D(\hat a)$, the drift of the exact Wigner equation~\eqref{eq:wigner_equation} is therefore the flow of Eq.~\eqref{eq:semiclassical_flow} at every $\aleph$. Only the diffusion and the third-order Kerr term depend on $\aleph$.

With the bare detuning held fixed instead, the Wigner drift is the flow at detuning $\Delta_{\rm bare}+\widetilde U/\aleph$, a shift of the same order in $1/\aleph$ as the diffusion. A scan of $\aleph$ at fixed $\Delta_{\rm bare}$ therefore moves the attractors, saddles, and basin boundaries of the flow while it changes the quantum noise. An $\aleph$ dependence measured along such a scan mixes quantum corrections with this change of the classical dynamics. For instance, for an activated rate $\propto\exp[-\aleph S(\Delta)]$ with a smooth action $S$, the shift changes the logarithm of the rate by $-\widetilde U\partial_\Delta S$, an amount of order one; at a crossover between competing escape paths, $S$ need not be smooth. At fixed $\aleph$, the two conventions describe the same Liouvillian with relabeled detuning.

The Kerr-shifted convention is therefore essential for the quantum-to-classical correspondence used in this work. It makes the scan of $b/s$ over $\aleph=5$, $7$, and $10$ (Sec.~\ref{subsec:two_photon_parity_bright_vacuum}) a scan at fixed classical dynamics, and $\aleph\to\infty$ the semiclassical limit of that fixed flow. At a given $\aleph$, it makes the quoted detuning the detuning of the exact Wigner drift. The mean-field attractors used as coherent anchors (Sec.~\ref{subsec:biased_kerr_kinetic_rerouting}) and the basins compared with the quantum branching probabilities (Appendix~\ref{app:biased_asymmetries}) therefore belong to the quantum dynamics at the same $\aleph$. Evaluated at $\Delta_{\rm bare}$, the flow would be displaced by $\widetilde U/\aleph=0.2\,\widetilde U$ at $\aleph=5$, which is $4\%$ of $\Delta$ at point $A$.

\subsection{Correlations and response}
\label{app:correlations_response}

For Hermitian inputs and readouts, a nonreal mode and its partner contribute $2\operatorname{Re}[e^{\lambda_kt}w_k(\hat O,\hat\rho_{\rm in})]$ to Eq.~\eqref{eq:one_shot_amplitude}. The same factorization governs stationary correlations and response through the quantum regression theorem~\cite{Lax1963,Gardiner2004}. For a source superoperator $\mathcal B$ and the connected source $\hat X_{\mathcal B}^\perp=\mathcal B(\hat\rho_{\rm ss})-\hat\rho_{\rm ss}\operatorname{Tr}[\mathcal B(\hat\rho_{\rm ss})]$, the one-sided transform $\int_0^\infty dt\,e^{-i\omega t}(\cdot)$ of $\operatorname{Tr}[\hat Ae^{\mathcal Lt}\hat X_{\mathcal B}^\perp]$ is
\begin{equation}
	\begin{aligned}
	C_{A\mathcal B}^{\rm conn,+}(\omega)
	&=\sum_{k>0}\frac{\mathcal R_k^{A\mathcal B}}{i\omega-\lambda_k}
	\;,\\
	\mathcal R_k^{A\mathcal B}&=D_k(\hat A)E_k\big(\mathcal B(\hat\rho_{\rm ss})\big)
	\;.
	\end{aligned}
	\label{eq:liouvillian_pole_expansion}\alsolabel{eq:residue_definition_general}
\end{equation}
The pole fixes frequency and linewidth, while the residue fixes amplitude and phase in the chosen source-readout channel; changing the source or readout can enhance, suppress, or phase shift a spectral feature without moving its pole~\cite{Scarlatella2019,Seibold2026,CamposVenuti2016}. Left multiplication, $\mathcal B(\hat\rho)=\hat B\hat\rho$, gives the connected correlation $\langle\hat A(t)\hat B(0)\rangle_{\rm ss}-\langle\hat A\rangle_{\rm ss}\langle\hat B\rangle_{\rm ss}$, and $-i[\hat B,\hat\rho_{\rm ss}]$ gives the retarded Hamiltonian response.

\subsection{Resolved spectral sectors}
\label{app:resolved_sectors}

A cluster $S$ of eigenvalues is resolved when a closed contour $\Gamma_S$ in the resolvent set encloses it and no other eigenvalue. This isolation is a spectral condition, distinct from the numerical conditioning of the cluster, which the resolvent norm and the Petermann factors below control, and from the experimental resolution of its individual poles. Its Riesz projector
\begin{equation}
	\mathcal P_S=\frac{1}{2\pi i}\oint_{\Gamma_S}dz\,(z-\mathcal L)^{-1}
	\label{eq:riesz_spectral_projector}
\end{equation}
reduces to $\sum_{k\in S}\hat r_k\operatorname{Tr}(\hat\ell_k^\dagger\,\cdot\,)$ for a diagonalizable cluster and retains the full generalized eigenspace at a defective point. The cluster contribution to a transient, $\operatorname{Tr}[\hat Oe^{\mathcal Lt}\mathcal P_S(\hat X)]$, is invariant under basis changes within the cluster but need not factorize into scalar modal weights. Near an internal degeneracy, individual eigenpairs can be ill conditioned, as measured by the Petermann factor $\|\hat r_k\|_{\rm HS}^2\|\hat\ell_k\|_{\rm HS}^2$~\cite{Petermann1979,Trefethen2005,Wiersig2023}, while $\mathcal P_S$ can remain well conditioned~\cite{Kato1995,Albert2014,Minganti2019,Pan2026}: $\|\mathcal P_S\|\le(|\Gamma_S|/2\pi)\max_{z\in\Gamma_S}\|(z-\mathcal L)^{-1}\|$, where $|\Gamma_S|$ is the length of the contour. For a non-normal $\mathcal L$, this resolvent norm, not the distance of the contour from the rest of the spectrum alone, controls the conditioning. For the slow sectors analyzed here a cluster-level treatment was not needed: their separation $\Gamma_{\rm rest}/\Gamma_{\rm sector}$ is at least $10$ where results are reported, and at $A$ and $B$ the individual slow eigenpairs have biorthogonal condition numbers below $7$ (Table~\ref{tab:endpoint}). We therefore use individual modes only while they are isolated, continuously trackable, and well conditioned, and otherwise the restricted propagator of the sector. The slowest decay rate ranks lifetimes, but the modal weights determine which modes dominate a given protocol~\cite{Mori2020,Mori2023,Shirai2024,Beato2026}. For a simple real mode in a Hermitian gauge, a regular zero of $\mathcal E_k(X,P)$ is a codimension-one contour; for a complex map, real and imaginary parts must vanish separately.

\section{Phase-space conventions and numerical procedures}
\label{app:phase_space_numerics}

\subsection{Phase-space symbols and mode gauges}
\label{app:phase_space_conventions}

With $\hat D(\alpha)=\exp(\alpha\hat a^\dagger-\alpha^*\hat a)$ and $\hat\Pi=\exp(i\pi\hat a^\dagger\hat a)$, we use the Wigner transform
\begin{equation}
	\begin{aligned}
	\mathcal W(\hat A;\alpha)&=\frac{2}{\pi}\operatorname{Tr}\!\left[\hat D^\dagger(\alpha)\hat A\hat D(\alpha)\hat\Pi\right]
	\;,\\
	\int d^2\alpha\,\mathcal W(\hat A;\alpha)&=\operatorname{Tr}(\hat A)
	\;,
	\end{aligned}
	\label{eq:app_wigner_transform}
\end{equation}
with $\alpha(X,P)=(X+iP)/\sqrt2$. Figures plot these symbols against the scaled quadratures $X/\sqrt\aleph$ and $P/\sqrt\aleph$. Since $\operatorname{Tr}(\hat r_k)=0$ for $k>0$, the right-mode symbol $W_k^R$ integrates to zero; it shows how mode $k$ changes the phase-space density. With the Husimi symbol $Q_{\hat A}(\alpha)=\pi^{-1}\langle\alpha|\hat A|\alpha\rangle$, the excitation map is $\mathcal E_k(X,P)=\pi Q_{\hat\ell_k^\dagger}(\alpha(X,P))$. The two panels of a mode thus represent different operators through different symbols, and only their supports, signs, and nodal geometry are compared. A readout detects mode $k$ through the overlap of its Weyl symbol with $W_k^R$.

For an isolated mode, we fix the reciprocal amplitude by
\begin{equation}
	\|\hat r_k\|_{\rm HS}=1
	\;,\qquad
	\operatorname{Tr}(\hat\ell_k^\dagger\hat r_k)=1
	\;.
	\label{eq:app_mode_amplitude_gauge}
\end{equation}
For a simple real eigenvalue, the matched eigenoperators are chosen Hermitian. The remaining common sign flips both $W_k^R$ and $\mathcal E_k$ but no modal weight; along a sweep we fix it by a positive Hilbert--Schmidt overlap with the previously tracked right eigenoperator. For a complex branch the same overlap is chosen real and positive, and figures show a stated quadrature or $|\mathcal E_k|$. Only the modulus has the same zeros as $\mathcal E_k$; $\operatorname{Re}(\mathcal E_k)$ or $\operatorname{Im}(\mathcal E_k)$ alone can also vanish elsewhere.

\subsection{Spectral diagnostics, parity adaptation, and mode tracking}
\label{app:kerr_cutoff_parity_selection}

For every retained eigenpair, we evaluate the residuals
\begin{equation}
	\begin{aligned}
	\varepsilon_k^R&=\frac{\|\mathcal L\hat r_k-\lambda_k\hat r_k\|_{\rm HS}}{\|\mathcal L\|\,\|\hat r_k\|_{\rm HS}}
	\;,\\
	\varepsilon_k^L&=\frac{\|\mathcal L^\dagger\hat\ell_k-\lambda_k^*\hat\ell_k\|_{\rm HS}}{\|\mathcal L\|\,\|\hat\ell_k\|_{\rm HS}}
	\;,
	\end{aligned}
	\label{eq:app_eigenpair_residuals}
\end{equation}
and the biorthogonality defect $\|\mathsf V_L^\dagger\mathsf V_R-\mathbb I\|_2$ of the matched retained eigenoperators. These residuals are normalized by $\|\mathcal L\|$, whereas Table~\ref{tab:endpoint} divides by $|\lambda_j|$, a much stricter measure for slow modes. Eigenvalues, residuals, overlaps, and maps must be stable under enlargement of the cutoff and changes of continuation step and solver tolerances. At $F=0$, a numerical eigensolver may return arbitrary mixtures within numerically degenerate eigenspaces; diagonalizing the projected parity action $(\mathsf Z_2)_{\mu\nu}=\operatorname{Tr}[\hat\ell_\mu^\dagger\mathcal Z_2(\hat r_\nu)]$ within each such cluster gives a parity-adapted basis, with the inverse-adjoint transformation applied to the left operators. Distinct eigenvalues are not mixed, since for $[\mathcal L,\mathcal Z_2]=0$ they already carry definite parity. At finite bias, an isolated branch is continued by the gauge-invariant score $|\operatorname{Tr}(\hat\ell_k^\dagger\hat r_j')\operatorname{Tr}(\hat\ell_j'^\dagger\hat r_k)|$ between consecutive parameter points; a resolved cluster is continued through its Riesz projector.

\subsection{Construction of reduced slow coordinates}
\label{app:slow_coordinate_construction}

For the three-coordinate reductions, the Hermitian trace-one slice $[\hat\rho_{\rm ss}+\operatorname{span}(\hat r_1,\hat r_2)]\cap\{\hat A=\hat A^\dagger\}$ is a real affine plane on which every operator has unit trace. The accessible coordinate range $\{(\operatorname{Tr}(\hat\ell_1^\dagger\hat\rho),\operatorname{Tr}(\hat\ell_2^\dagger\hat\rho)):\hat\rho\ge0\}$ is convex; its boundary is sampled by the top eigenvectors of $\cos(\theta)\hat\ell_1+\sin(\theta)\hat\ell_2$ on a grid of directions $\theta$. The positive section $\{(x,y):\hat\rho_{\rm ss}+x\hat r_1+y\hat r_2\ge0\}$ is a convex subset of this range containing $\hat\rho_{\rm ss}$, since the two coordinate charts coincide by biorthogonality; along each direction $\theta$ its boundary is located by bisection for the largest $t$ with $\lambda_{\min}\{\hat\rho_{\rm ss}+t[\cos(\theta)\hat r_1+\sin(\theta)\hat r_2]\}\ge0$. Following the simplex construction for metastable manifolds and robust Perron clusters~\cite{Macieszczak2021,Deuflhard2005,Roeblitz2013}, the representatives $\hat\sigma_c,\hat\sigma_\pm$ are the vertices of a triangle inscribed in the convex hull of the chosen region and containing the stationary point. The region is either the accessible range (left-accessible representatives) or its intersection with the positive section (positive representatives). Among the $48$ largest such triangles with at least $80\%$ of the maximal area, we prefer positive representatives, then the smallest classicality defect~\cite{Macieszczak2021}, then continuity along a sweep. The representatives are labeled by their phase-space localization. The construction uses only the geometry of the slow manifold, not a kinetic model.

With the expansion of Sec.~\ref{subsec:two_photon_parity_bright_vacuum}, every state in the retained affine subspace satisfies $\hat\rho=\sum_jp_j(\hat\rho)\hat\sigma_j$ with $\sum_jp_j(\hat\rho)=\operatorname{Tr}(\hat\rho)$. For fixed representatives, the duals and coordinates are invariant under mode-gauge transformations and basis changes within the slow subspace, but they depend on the representatives. The projected generator $Q_{ji}=\operatorname{Tr}[\hat\eta_j^\dagger\mathcal L(\hat\sigma_i)]$ has columns labeled by the initial and rows by the final representative and vanishing column sums. We retain it exactly and assign a stochastic interpretation only where it already satisfies the admissibility conditions below, without replacing it by a nearest stochastic generator~\cite{Macieszczak2021}. It differs from a switching matrix inferred from trajectory statistics, as used for coexisting limit cycles~\cite{Nowoczyn2026}. The preparation regions of Fig.~\ref{fig:biased_kerr_membership_rerouting} use $p_j(\alpha)=p_j(|\alpha\rangle\langle\alpha|)$; their color saturation is the margin $p_{(1)}(\alpha)-p_{(2)}(\alpha)$ between the two largest coordinates.

\paragraph*{Symmetric three-state generator.}\label{app:symmetric_three_state}
In the ordered basis $(c,+,-)$, the generator of Eq.~\eqref{eq:cat_symmetric_three_state_generator} has the right eigenvectors $\mathbf v_o\propto(0,1,-1)^{\mathsf T}$ and $\mathbf v_e\propto(1,-1/2,-1/2)^{\mathsf T}$, with the eigenvalues of Eq.~\eqref{eq:cat_symmetric_nonstationary_rates}. The matched even left eigenvector is $\mathbf u_e^{\mathsf T}=(a,-b,-b)/(a+b)=(1-\pi_c,-\pi_c,-\pi_c)$ by Eq.~\eqref{eq:cat_symmetric_stationary_weights}, so for fixed representatives it carries no information on the rates beyond $\pi_c$.

\subsection{Positivity, stochastic admissibility, and delay choice}
\label{app:slow_coordinate_checks}

The reduction must satisfy three separate requirements~\cite{Macieszczak2021}. A representative is a physical state only if it is positive semidefinite (Hermiticity and unit trace hold by construction). The coordinates are probabilities only on state sets where all $p_i(\hat\rho)\geq0$. The projected generator is a Markov generator only if it is real, column conserving, and nonnegative off the diagonal. Representative positivity implies neither of the other two. We monitor
\begin{equation}
	\begin{aligned}
	\varepsilon_\sigma&=\max_i\max\big(0,-\lambda_{\min}(\hat\sigma_i)\big)
	\;,\\
	\varepsilon_p&=\max_{i,\,\hat\rho\in\mathcal S_{\rm samp}}\max\big(0,-p_i(\hat\rho)\big)
	\;,
	\end{aligned}
	\label{eq:app_representative_coordinate_violations}
\end{equation}
on a sampled state set $\mathcal S_{\rm samp}$, together with the largest imaginary part, column-sum defect, and negative off-diagonal entry of $\mathbf Q$. We compare these defects with the numerical tolerance and check them under changes of cutoff, continuation step, and representative selection; defects above tolerance are reported. A sampled check does not prove global coordinate positivity. Where the tests fail, the affine reduction remains well defined but does not establish a stochastic model of physical states. Even where they hold, identifying the entries with first-passage rates, or coordinate boundaries with committors, requires independent dynamical validation.

A delay-design benchmark uses three preparations, Gaussian shot noise from full $N=80$ evolution, calibration drift, and $2\%$ pole errors. For shot noise alone, $t_*$ lies within $5\%$ of the best single delay. With all errors at $6\times10^4$ shots, the optimum moves to $0.575t_*$ with a $9\%$ error penalty at $t_*$; the kernel maximum $t_*$ is thus a convenient, not an optimal, delay once calibration and pole errors are included. The calibration must be taken on the slow plateau: with noiseless inputs, plateau calibration gives a relative reconstruction error of $9.9\times10^{-8}$ at $t_*$, immediate calibration $5.3\%$, because an early calibration changes the inferred preparation basis.

\subsection{Detuning scans of the linearly and parametrically driven resonators}
\label{app:detuning_scans}

\begin{figure}[t]
	\centering
	\includegraphics[width=\linewidth]{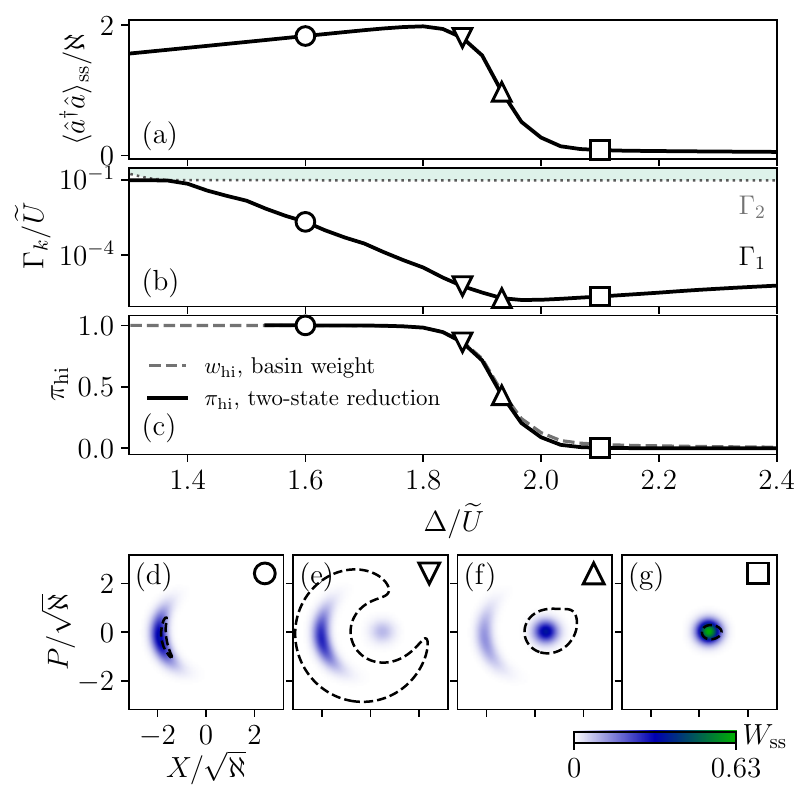}
	\caption{\textit{Switching mode of the linearly driven resonator.} (a) Normalized stationary occupation $\langle\hat n\rangle_{\rm ss}/\aleph$. (b) Switching rate $\Gamma_1=\Gamma_{\rm sw}$ and next decay rate $\Gamma_2$ (dotted), with faster rates in the shaded region. (c) Stationary bright-state coordinate $\pi_{\rm hi}$ for $\Gamma_2/\Gamma_1\geq10$ (solid) and stationary Husimi weight $w_{\rm hi}$ integrated over the classical bright basin (dashed). We set $w_{\rm hi}=1$ below the bistability threshold $\Delta/\widetilde U\simeq1.19$. The two weights agree within $0.02$ up to $\Delta/\widetilde U=1.95$; at larger detunings, the Husimi weight exceeds $\pi_{\rm hi}$ by up to $0.04$ due to the tails of the dim-state distribution. The switching mode separates from the faster spectrum near $\Delta/\widetilde U=1.5$. (d--g) Stationary Wigner functions and excitation-zero contours (dashed) at $\Delta/\widetilde U=1.60$, $1.87$, $1.93$, and $2.10$. At $\Delta/\widetilde U=1.6$, the reduced representatives have a negative eigenvalue of $-0.05$. Parameters: $\widetilde F/\widetilde U=0.5$, $\kappa/\widetilde U=0.1$, $G=0$, and $\aleph=5$.}
	\label{fig:kerr_bistable_switching_tomography}
\end{figure}

Figure~\ref{fig:kerr_bistable_switching_tomography} follows the switching mode of Sec.~\ref{subsec:one_photon_switching_basin} across the detuning-induced crossover from bright to dim stationary occupation. The switching rate becomes more than four orders of magnitude smaller than the next decay rate, while the stationary bright-state coordinate $\pi_{\rm hi}$ decreases from nearly one to nearly zero. The right-mode Wigner symbol changes little, with normalized overlap $0.94$ between $\Delta/\widetilde U=1.6$ and $1.9$, whereas the excitation-zero contour moves substantially, as required by $\mathcal E_{\rm sw}\propto p_{\rm hi}-\pi_{\rm hi}$. As $\pi_{\rm hi}$ decreases, the contour passes through preparations far from the stationary state that nevertheless suppress the switching mode.

\begin{figure}[t]
	\centering
	\includegraphics[width=\linewidth]{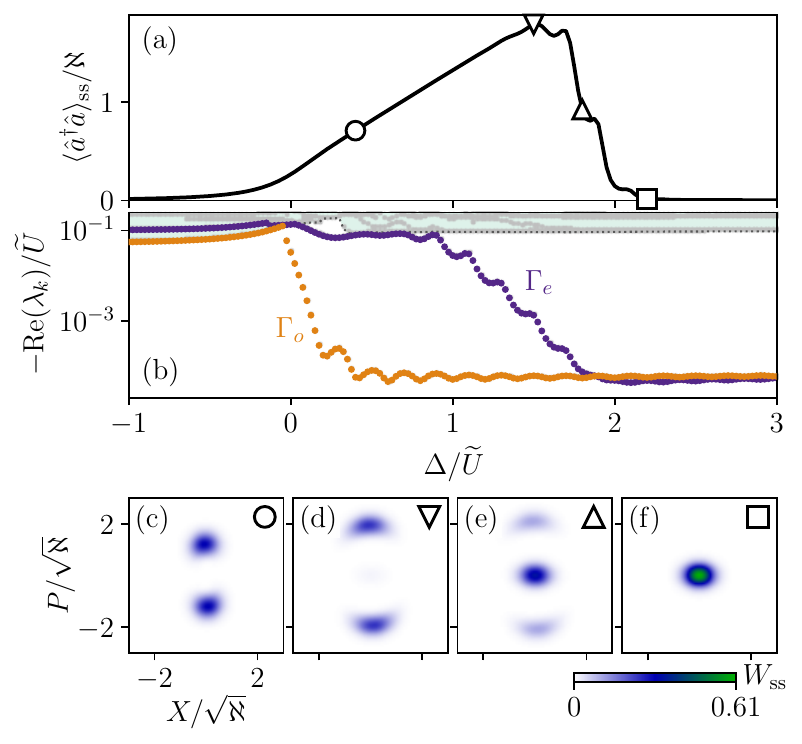}
	\caption{\textit{Detuning scan of the parametrically driven Kerr resonator.} (a) Normalized stationary occupation $\langle\hat n\rangle_{\rm ss}/\aleph$. (b) Low-lying decay rates, highlighting the slowest odd ($\Gamma_o$) and even ($\Gamma_e$) modes. The shaded region contains the faster decay rates; $\lambda_0=0$ is omitted. (c--f) Stationary Wigner functions at $\Delta/\widetilde U=0.4$, $1.5$, $1.8$, and $2.2$, showing the crossover from outer-lobe to central occupation. The last three detunings correspond to Fig.~\ref{fig:cat_transition_tomography}. Parameters: $G/\widetilde U=0.5$, $F=0$, $\kappa/\widetilde U=0.1$, and $\aleph=5$.}
	\label{fig:cat_detuning_scan}
\end{figure}

\begin{figure}[t]
	\centering
	\includegraphics[width=\linewidth]{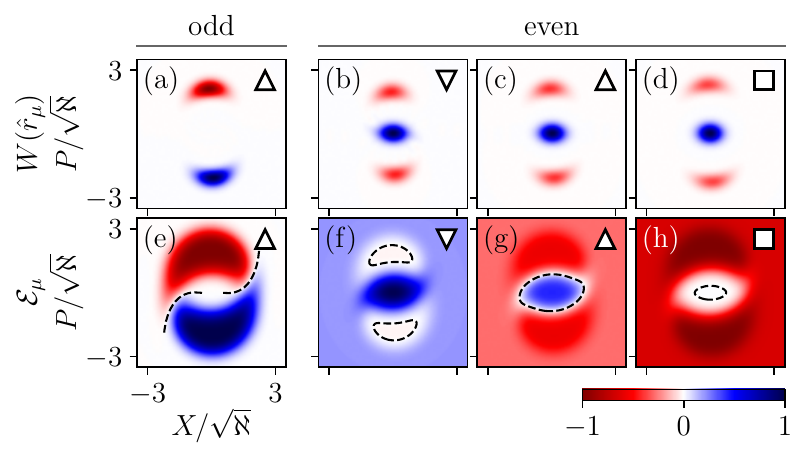}
	\caption{\textit{Slow modes across the crossover to central stationary occupation.} Top row: right-mode Wigner symbols $W_\mu^R(X,P)$ normalized to their maximum modulus. Bottom row: excitation maps $\mathcal E_\mu(X,P)$ with dashed zero contours. (a,e) Odd mode at $\Delta/\widetilde U=1.8$, with $\mathcal E_o$ normalized by its value at the lower lobe and proportional to $p_--p_+$. (b--d,f--h) Even mode at $\Delta/\widetilde U=1.5$, $1.8$, and $2.2$, with $\pi_c=0.02$, $0.57$, and $1.00$. The excitation maps are normalized by the difference between their values at the central and outer attractors and are proportional to $p_c-\pi_c$, which equals $1-\pi_c$ and $-\pi_c$ at the central and outer representatives, respectively. Signs are chosen positive at the lower lobe (odd) and center (even); blue (red) denotes positive (negative) values. Parameters as in Fig.~\ref{fig:cat_detuning_scan}.}
	\label{fig:cat_transition_tomography}
\end{figure}

Figures~\ref{fig:cat_detuning_scan} and~\ref{fig:cat_transition_tomography} show the parametrically driven resonator at $G/\widetilde U=0.5$, where the crossover from outer-lobe to central stationary occupation lies within $\Delta/\widetilde U\leq2.2$. At the operating point $(\Delta,G)/\widetilde U=(5,0.4)$ of Secs.~\ref{subsec:biased_kerr_kinetic_rerouting}--\ref{sec:v11identification}, the center is already nearly absorbing ($\pi_c\simeq1$), so the crossover is not visible there. The odd rate $\Gamma_o$ decreases by three orders of magnitude near $\Delta/\widetilde U=0$ and then remains near $6\times10^{-5}\,\widetilde U$. The even rate decreases between $\Delta/\widetilde U\simeq0.9$ and $1.8$ and becomes smaller than $\Gamma_o$ beyond $\Delta/\widetilde U\simeq1.85$, while the outer lobes remain metastable. The odd mode changes little across the crossover. With $\mathbf v_e=(1,-1/2,-1/2)^{\mathsf T}$, the even left eigenvector is $\mathbf u_e^{\mathsf T}=(1-\pi_c,-\pi_c,-\pi_c)$, so $\mathcal E_e(X,P)\propto p_c(X,P)-\pi_c$ and the largest even-mode excitation amplitude moves from the central to the outer representatives as $\pi_c$ crosses $1/2$; the right eigenoperator continues to describe the same population contrast.

\subsection{Biased slow sector: coordinate checks and bias reversal}
\label{app:biased_asymmetries}

At zero bias, the representatives of the left-accessible construction keep the negativity $\varepsilon_\sigma\simeq0.0196$ when the cutoff is raised from $N=60$ to $N=80$; these representatives define affine coordinates but are not physical density operators. Selecting positive representatives instead does not ensure nonnegative coordinates of coherent preparations, so spectral separation alone does not establish a stochastic three-state model over the displayed range. The center-escape entries $\Gamma_{\pm\leftarrow c}$ of Eq.~\eqref{eq:biased_three_state_generator_pm} remain below numerical resolution over the sweep: at the validated biased points the central representative is numerically indistinguishable from the stationary state, so negligible escape is a property of this construction, not an independently resolved rate. The zero crossings of $\mathcal R_\pm$ are properties of the projected representation, not topological boundaries of the semiclassical flow~\cite{Seibold2026}.

Bias reversal, $\mathcal Z_2\mathcal L_F\mathcal Z_2^{-1}=\mathcal L_{-F}$, maps negative-bias quantities onto positive-bias ones by $(X,P)\mapsto(-X,-P)$ and $+\leftrightarrow-$, up to the tracked slow-subspace gauge. Analytically continued simple poles are therefore even in $F$, whereas the difference of the two inter-lobe branching probabilities is odd for parity-covariant coordinates and can have a linear term if the construction is differentiable.

\paragraph*{Quasienergy asymmetries under bias.}
For real $F$ and $G$, complex conjugation corresponds to the phase-space reflection $(X,P)\mapsto(X,-P)$ and preserves the quasienergy of the conservative fixed points at $\kappa=0$. It exchanges the outer lobes and leaves each saddle invariant. A transverse drive (real $F$) therefore preserves the outer-lobe quasienergy degeneracy while splitting the saddle quasienergies by $1.716\,\widetilde U$ at $|\widetilde F|/\widetilde U=0.2$. An in-phase drive (imaginary $F$), by contrast, splits the outer-lobe quasienergies by $1.859\,\widetilde U$ while preserving the saddle degeneracy. Complex conjugation reverses the Hamiltonian contribution to the Lindblad dynamics but leaves the dissipator invariant. The reflection therefore does not enforce equal transition rates in the damped resonator, and the quasienergy asymmetries alone do not determine the preferred exits.

\paragraph*{Prepared basis, representatives, and spread of the branching probabilities.}
The main text uses the prepared basis, $\mathbf Q'=P_0^{-1}\mathbf QP_0$, whose three coherent preparations are physical states (Sec.~\ref{subsec:biased_kerr_kinetic_rerouting}). The most negative coordinate on the sampled coherent-state domain in this basis is $-1.3\times10^{-3}$. The left-accessible representatives of Fig.~\ref{fig:biased_kerr_directional_branching}(a--c), tracked from $\widetilde F=0$, serve as a check. Table~\ref{tab:routing_spread} compares the branching probabilities at three biases from four procedures. The first is the prepared basis with two perturbed preparations: displaced thermal states with mean thermal photon number $0.05$, and coherent states moved to the nearest point of a grid of spacing $0.25$ in $X/\sqrt\aleph$ and $P/\sqrt\aleph$. The second uses the left-accessible representatives and positive representatives inscribed in the states of the slow slice. The third is the splitting probability of Fig.~\ref{fig:biased_kerr_directional_branching}(e), and the fourth is first-exit counting in quantum trajectories. All procedures give center capture as the preferred exit of both lobes at $\widetilde F/\widetilde U=0.1$ and $B_{+\to-}>1/2$ at $0.2$. At $0.3$ both lobes prefer switching in every procedure except the splitting probabilities with absorption rates $\gamma/\widetilde U\le0.01$. At $0.2$, $B_{-\to+}$ ranges from $0.44$ to $0.52$ over the reductions and the splitting band, so the sign of $\mathcal R_-$ is not resolved there. The zeros of $\mathcal R_+$ lie in $[0.158,0.163]$ and those of $\mathcal R_-$ in $[0.197,0.213]$ across the prepared basis, both representative rules, and the splitting band; the prepared basis gives $0.1616$ and $0.2088$, the left-accessible representatives $0.1614$ and $0.2050$. The trajectory estimates are exploratory: they are not converged in time, depend on the classifier, and, unlike the ensemble quantities used elsewhere in this paper, can depend on the unravelling; at $\widetilde F/\widetilde U=0.2$ their $B_{-\to+}=0.30$ lies outside the range of the other procedures. Right censoring is substantial at this point (Table~\ref{tab:routing_spread}); the quoted counting estimates do not establish asymptotic splitting probabilities. A converged competing-risk analysis would need to account for the surviving trajectories and their possible subsequent destinations.

The flip-rate dip of Fig.~\ref{fig:biased_kerr_directional_branching}(d) depends on the basis below the resolution of the reduction. In the prepared basis, $\Gamma_{+\leftarrow-}$ is a smooth function of the bias with zeros at $\widetilde F/\widetilde U=0.023$ and $0.045$ (linear interpolation on the grid of step $0.0025$) and negative values in between, with minimum $-2.5\times10^{-7}\,\widetilde U$ at $0.0325$; relative to $\Gamma_{c\leftarrow-}$ it reaches at most $-8.8\times10^{-4}$. Fig.~\ref{fig:biased_kerr_directional_branching}(d) therefore draws its modulus over the whole range, dotted where it is negative; the perturbed preparations give minima of $-3.9\times10^{-7}\,\widetilde U$ and $-1.9\times10^{-7}\,\widetilde U$ on slightly different intervals. With the left-accessible representatives the entry stays positive, with minimum $1.7\times10^{-9}\,\widetilde U$ at the same bias, and with positive representatives it vanishes to round-off for $\widetilde F/\widetilde U\ge0.01$. Each value is converged in the cutoff ($N=60$ and $80$ agree to $2\times10^{-12}\,\widetilde U$), but the spread over bases and preparations, $3.8\times10^{-7}\,\widetilde U$, exceeds the entry, so its sign is not robust under these changes. The splitting probability $q(-)$ has its minimum at $\widetilde F/\widetilde U=0.030$ for 18 of 21 settings of absorption rate and region radius ($0.035$--$0.040$ for the other three), and lies a factor $1.2$ to $7.9$ below its zero-bias value, depending on the setting.

\begin{table*}[t]
	\centering
	\caption{Branching probabilities $B_{+\to-}/B_{-\to+}$ at three biases from different procedures, $N=60$ unless stated. Prepared basis and representatives agree at $N=60$ and $80$. Splitting probabilities: ranges over absorption rates $\gamma$ and region radii $r$ (in units of $\sqrt\aleph$). Trajectories: first-exit fractions among the trajectories that leave their start within $\widetilde Ut\le2\times10^4$, classified by phase-space core regions and therefore dependent on the classifier; $200$ per start, $68\%$ Wilson intervals of half-width at most $0.036$. Trajectories without an exit (censored) are excluded from the denominator; their fraction among the $-$ starts is given in parentheses.}
	\label{tab:routing_spread}
	\footnotesize
	\begin{tabular}{lccc}
		\toprule
		Procedure & $\widetilde F/\widetilde U=0.1$ & $0.2$ & $0.3$ \\
		\midrule
		Prepared basis & $0.226/0.041$ & $0.661/0.450$ & $0.911/0.863$ \\
		Perturbed preparations & $0.225/0.038$--$0.043$ & $0.660/0.441$--$0.455$ & $0.908$--$0.910/0.858$--$0.869$ \\
		Left-accessible repr. & $0.225/0.047$ & $0.663/0.471$ & $0.914/0.874$ \\
		Positive repr. & $0.228/0.000$ & $0.662/0.452$ & $0.915/0.781$ \\
		Splitting, $\gamma/\widetilde U\in[0.1,1]$, $r\in\{0.5,0.75\}$ & $0.222$--$0.250/0.046$--$0.066$ & $0.655$--$0.679/0.444$--$0.520$ & $0.907$--$0.918/0.847$--$0.901$ \\
		Splitting, also $r=1$ & $0.222$--$0.307/0.046$--$0.113$ & $0.655$--$0.712/0.444$--$0.573$ & $0.907$--$0.929/0.847$--$0.923$ \\
		Splitting, $\gamma/\widetilde U\in[10^{-3},1]$, all $r$ & $0.138$--$0.307/0.013$--$0.113$ & $0.513$--$0.712/0.101$--$0.573$ & $0.812$--$0.929/0.307$--$0.923$ \\
		Trajectories (finite window) & $0.255/0.032$ ($6\%$) & $0.705/0.302$ ($15.5\%$) & $0.920/0.759$ ($15\%$) \\
		\bottomrule
	\end{tabular}
\end{table*}

\begin{figure}[t]
	\centering
	\includegraphics[width=0.94\linewidth]{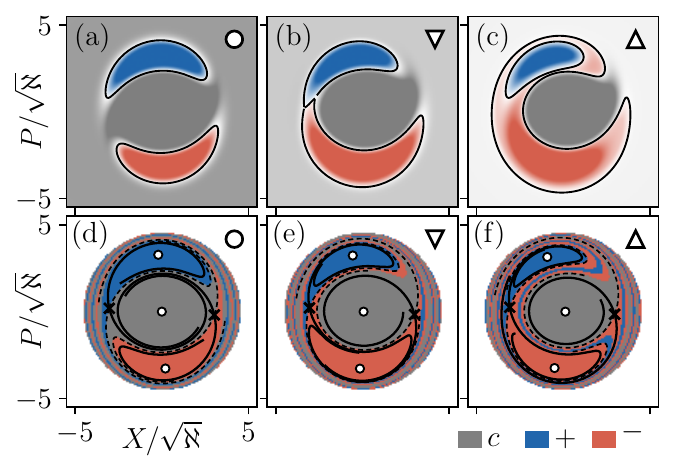}
	\caption{\textit{Preparation regions and mean-field basin connectivity.} (a--c) Dominant representative coordinate $p_j(\alpha)$ at $\widetilde F/\widetilde U=0$, $0.2$, and $0.4$. Color saturation denotes the difference between the two largest coordinates; black lines mark their equality. These regions depend on the representative construction and are not deterministic basins. The coherent-state truncation loss is below $2\times10^{-9}$ at $N=230$. (d--f) Mean-field basins of $c$ (gray), $+$ (blue), and $-$ (red), with unstable (solid) and stable (dashed) saddle manifolds. Unresolved interleaved basin bands beyond a scaled radius of approximately $4.5$ are omitted. Parameters as in Fig.~\ref{fig:biased_kerr_directional_branching}.}
	\label{fig:biased_kerr_membership_rerouting}
\end{figure}

\paragraph*{Preparation regions and mean-field basins.}
Figure~\ref{fig:biased_kerr_membership_rerouting}(a--c) partitions coherent preparations according to their largest representative coordinate $p_j(\alpha)=p_j(|\alpha\rangle\langle\alpha|)$. The adjacency changes from a star centered on $c$ to a chain $c$--$(-)$--$(+)$; at $\widetilde F/\widetilde U=0.2$, the $+$ region no longer borders the central region. This rearrangement accompanies the change in the preferred exit of the $+$ lobe, but the adjacency of preparation regions does not determine the subsequent transition probabilities. The mean-field basins provide a classical comparison [Fig.~\ref{fig:biased_kerr_membership_rerouting}(d--f)]. At $\widetilde F/\widetilde U=0.118$, a classical saddle connection leaves the $+$ basin bordering a saddle whose unstable branches connect the two outer lobes. This connectivity is consistent with the direction of the first quantum crossover, which occurs near $0.16$ at $\aleph=5$. The second quantum crossover near $0.21$ has no corresponding classical reconnection, the subsequent classical reconnections at $0.249$ and $0.331$ do not reverse the preferred quantum exits, and no reconnection accompanies the weak-bias switching minimum.

\subsection{Temporal kernel, finite return, and bias scan}
\label{app:rc_kernel_checks}

Starting in \(+\), Eq.~\eqref{eq:rc_kernel} gives \(p_-(t)=\pi_-[1-h(t)]+Q_{-,+}g(t)\); it overshoots \(\pi_-\) exactly when \(Q_{-,+}>\Gamma_2\pi_-\), and for an absorbing center it peaks at \(t_*\) with height \(Q_{-,+}g(t_*)\).

For nonnegative normalized stationary populations with $\pi_c>0$ and positive total lobe exit rates, the finite-return generalization of Eq.~\eqref{eq:rc_absorbing_bound} follows as below. Let \(u\) and \(v\) be the total exit rates of \(+\) and \(-\), \(x=B_{+\to-}\), \(y=B_{-\to+}\), \(a_\pm=\Gamma_{\pm\leftarrow c}\) the escape rates from the center, \(T=\Gamma_1+\Gamma_2\), and \(D=\Gamma_1\Gamma_2\), and define
\begin{equation}
	A=1+\frac{\pi_+}{\pi_c}(1-x)\;,
	\qquad
	C=1+\frac{\pi_-}{\pi_c}(1-y)\;,
	\label{eq:rc_finite_weight_AC}
\end{equation}
both at least one. The matrix-tree theorem gives \(uv(1-xy)=\pi_cD\). Stationarity gives \(\pi_ca_+=\pi_+u-\pi_-yv\) and \(\pi_ca_-=\pi_-v-\pi_+xu\), and with these the trace of the generator becomes \(T=Au+Cv\). Since \((Au+Cv)^2\ge4ACuv\), positive rates \(u\) and \(v\) exist only if
\begin{equation}
	T^2(1-xy)\ge4\pi_cDAC\;.
	\label{eq:rc_finite_weight_bound}
\end{equation}
This condition is necessary but not sufficient: a Markov generator also requires \(a_\pm\ge0\), that is, \(\pi_+u\ge\pi_-yv\) and \(\pi_-v\ge\pi_+xu\), for one of the roots \((u,v)\). Because \(A,C\ge1\), every Markov generator satisfying these assumptions obeys
\begin{equation}
	B_{+\to-}B_{-\to+}\le1-\frac{4\pi_c\Gamma_1\Gamma_2}{(\Gamma_1+\Gamma_2)^2}\;,
	\label{eq:rc_finite_weight_corollary}
\end{equation}
which reduces to Eq.~\eqref{eq:rc_absorbing_bound} for \(\pi_c\to1\). For $\pi_c>0$ and $D>0$ a feasible pair has $xy<1$. Put $Z=\pi_cD/(1-xy)$. The two candidate root pairs are
\begin{align}
	Au&=\frac{T\pm\sqrt{T^2-4ACZ}}{2},&
	v&=Z/u .
\end{align}
With $a_\pm$ fixed by stationarity, a pair yields a Markov generator when $a_+,a_-\ge0$. Its columns in order $(c,+,-)$ are
\begin{align}
	&(-a_+-a_-,a_+,a_-)^{\mathsf T},\notag\\
	&((1-x)u,-u,xu)^{\mathsf T},\quad
	((1-y)v,yv,-v)^{\mathsf T}.
\end{align}
This generator has the required trace and stationary weights by construction. Since the adjugate of a rank-two generator is proportional to $\pi\mathbf 1^{\mathsf T}$, the center cofactor $uv(1-xy)=\pi_cD$ fixes the product of the nonzero eigenvalues. This proves sufficiency.

Equality in Eq.~\eqref{eq:rc_finite_weight_bound} holds for a feasible generator if and only if \(Au=Cv\). The parity-symmetric generator, with \(A=C\) and \(u=v\), saturates this discriminant bound for every admissible \(\pi_c>0\).

For distinct matched real poles, let $P_1$ denote the spectral projector of $Q$ associated with $-\Gamma_1$. Then $Q=-\Gamma_2(I-\Pi)+(\Gamma_2-\Gamma_1)P_1$. For identical stationary coordinates, $Q_A-Q_B=(\Gamma_2-\Gamma_1)(P_{1,A}-P_{1,B})$, making explicit that the common-kernel contrast probes the difference between the slow spectral projectors in the chosen coordinate frames.

\paragraph*{Checks at the reference bias.}
The \(N=80\) calculation gives \(\widetilde U t_*=2174.6\), that is, \(t_*=217.5/\kappa\). Enlarging the cutoff from \(N=60\) changes the reduced generator by \(6.9\times10^{-9}\) relative to its largest entry. The stationary lobe weights are below numerical resolution, and the absorbing-center bound at this bias is \(0.8114\). Direct matrix exponentiation of the reduced generator reproduces the peak height \(0.5482\) of \(p_-(t)\) after preparing \(+\). For this generator, \(B_{+\to-}=0.661\), and the calibrated quadrature undershoots its stationary value.

\paragraph*{Matched slow poles.}
On the branch of Fig.~\ref{fig:rc_kernel}(b), \((\Delta,G)\) is solved by Newton iteration on \(\ln(\Gamma_1)\) and \(\ln(\Gamma_2)\) with continuation in \(\theta\) at \(N=110\) and checked at \(N=140\). On the separately refitted branch, both rates were matched to about \(2\times10^{-11}\) relative, and the projected stationary coordinates equal \((1,0,0)\) to \(4\times10^{-14}\). A solution at \(N=80\), checked at \(N=110\), agrees to \(5\times10^{-9}\) for \(\theta\le0.45\); at \(\theta=0.5\), where the lobes hold about \(48\) photons, this accuracy is lost. The spectral separation, including the complex fast modes, stays near \(35\), and the reduced generators are real, column conserving, and nonnegative off the diagonal.
\(B_{-\to+}\) stays below \(1/2\) on the whole branch, falling from \(0.450\) to \(0.060\).
At \(\theta=0\), varying \(|\widetilde F|/\widetilde U\) from \(0.15\) to \(0.25\) with \((\Delta,G)\) re-solved changes \(B_{+\to-}\) only from \(0.70\) to \(0.63\) and leaves \(B_{-\to+}\) near \(0.45\); at \(|\widetilde F|/\widetilde U=0.15\) the reduced generator fails the Markov test by \(8\times10^{-12}\,\widetilde U\), above the numerical floor. Over the tested intervals, the branching probabilities thus vary much less with the drive amplitude than with its phase.

\begin{figure*}[t]
	\centering
	\includegraphics[width=0.86\textwidth]{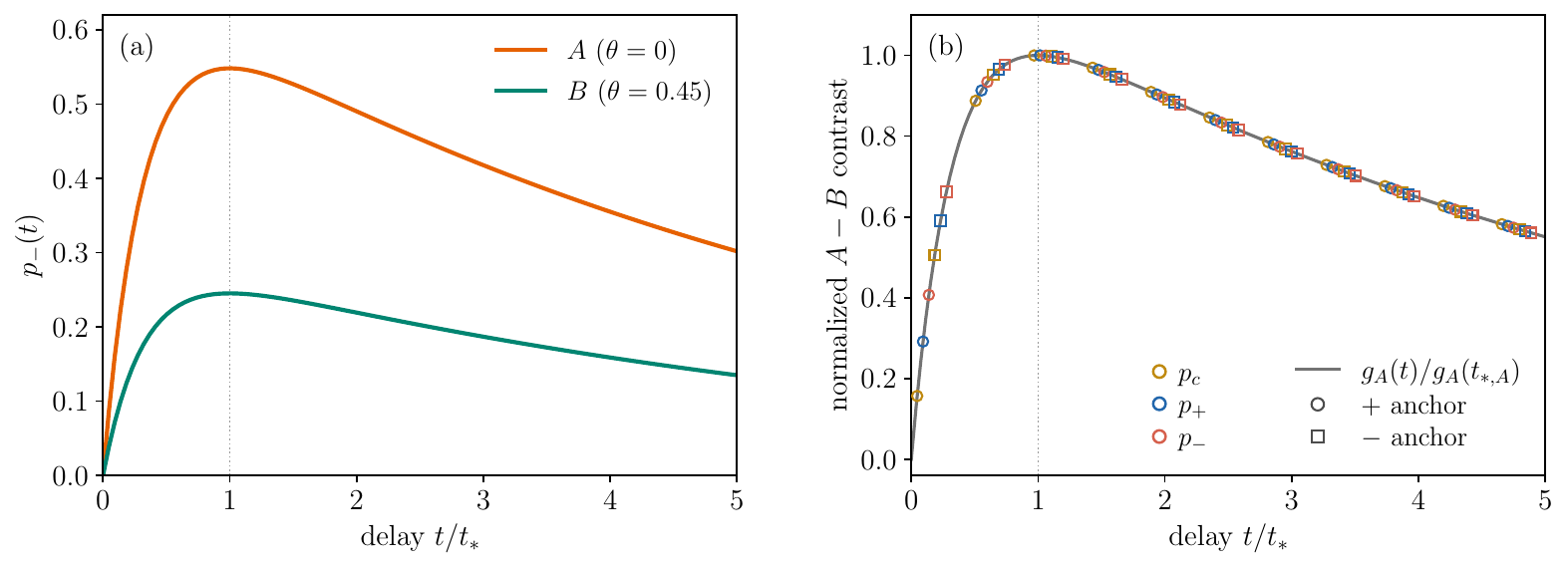}
	\caption{\textit{Calibrated population transients and common-kernel collapse at $N=80$.} (a) Opposite-lobe coordinates for the $+$ anchor from full Lindblad propagation with separate $A$ and $B$ readout calibrations. Near $t_*$ these reach approximately $0.55$ and $0.25$, respectively. (b) The three calibrated $A-B$ coordinate differences (gold $c$, blue $+$, red $-$) for each lobe anchor. Each is divided by its signed projected common-kernel prediction at $t_*$ for that coordinate and preparation. The gray curve is $g_A(t)/g_A(t_{*,A})$. Vertical lines mark $t/t_*=1$.}
	\label{fig:v16_kernel_collapse}
\end{figure*}

The $N=80$ trajectory record used in Figs.~\ref{fig:v11identifiability}(a,b) and \ref{fig:v16_kernel_collapse} comes from full Lindblad propagation at $\theta=0$ with $(\Delta,G)/\widetilde U=(5,0.4)$ and $\theta=0.45$ with $(8.91951842234025,0.409880038986779)$. The mean-field lobes hold $27.0$ and $46.5$ photons, and the slowest complex fast pair moves from $-0.050\pm4.78i$ to $-0.050\pm8.71i$ in units of $\widetilde U$. The relative slow-rate mismatches in this run are below $1.3\times10^{-11}$ (below $2.2\times10^{-11}$ in the exported projected generators used for the curves). The three anchor readout pairs at the plateau origins $\widetilde Ut_0=250.0$ and $245.3$ define separate calibrations, in which independent stationary readouts give lobe coordinates of magnitude at most $1.4\times10^{-9}$.

For the common-kernel comparison [Fig.~\ref{fig:v16_kernel_collapse}(b)], samples at the same delay index (delays agreeing to $5.2\times10^{-7}/\widetilde U$) are paired and $g_A(t)(Q_A-Q_B)p_0$ is subtracted from the calibrated coordinate difference. Relative to the largest predicted contrast of each preparation, the maximum residual is $4.1\times10^{-9}$ for the $+$ and $5.0\times10^{-8}$ for the $-$ anchor, and the center-prepared contrast is $4\times10^{-9}$ of the $+$ contrast. Evaluating each prediction with its own poles and stationary coordinates, which adds the term $[1-h(t)](\pi_A-\pi_B)$, leaves absolute residuals below $1.4\times10^{-9}$. The raw readout differences, unlike the coordinate differences, reflect the different lobe positions and do not follow $g(t)$.

With $10^4$ shots per setting, a pole error of $1\%$, and $400$ simulated realizations, the witness at $t_*$ gives $W_+=+0.267\pm0.009$ and $-0.358\pm0.008$ at $A,B$, respectively, compared with the projected-generator predictions $+0.2668$ and $-0.3585$. Every realization gives the correct sign. Under the noise assumptions of this benchmark, the estimated counts for resolving the sign at $2\sigma$ are $42$ and $19$ shots per setting.

\begin{figure*}[t]
	\centering
	\includegraphics[width=\textwidth]{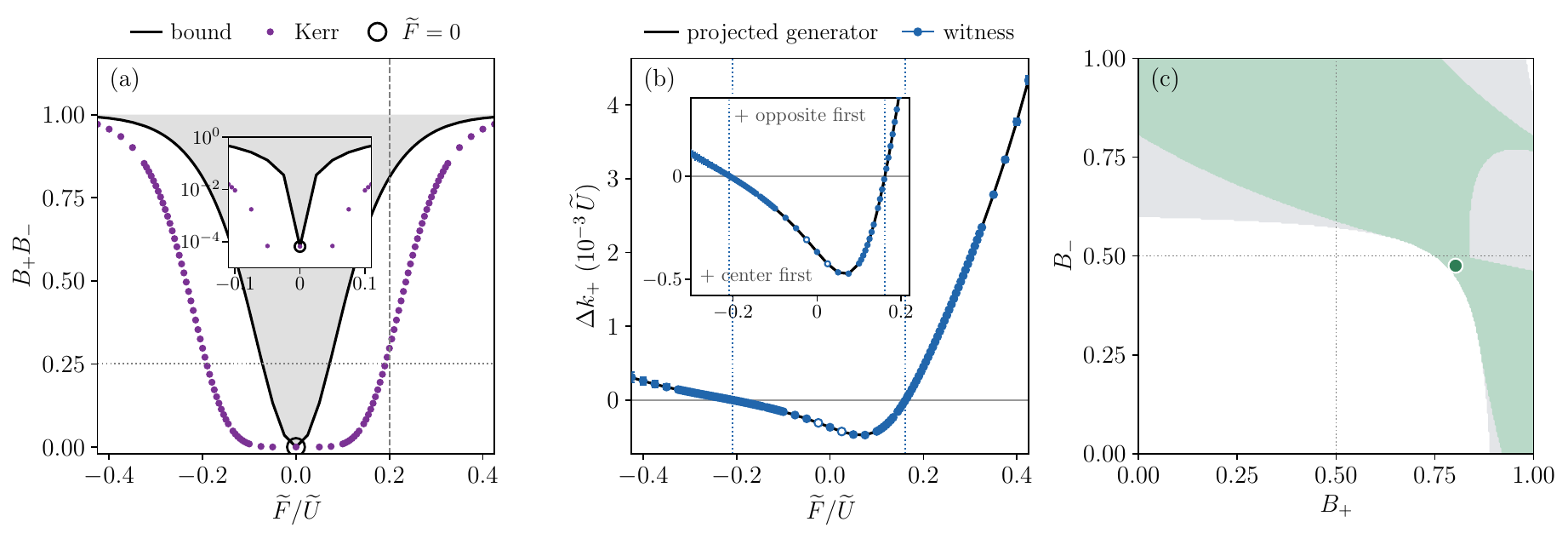}
	\caption{\textit{Spectral constraints and the directional witness.}
		(a) Absorbing-center bound of Eq.~\eqref{eq:rc_absorbing_bound} (black line, gray excluded region) and projected Kerr branching products (purple points) across the real-bias scan at $N=60$. The circle at zero bias marks parity saturation. The horizontal dotted line marks $1/4$, the vertical dashed line marks the reference bias, and the inset uses a logarithmic scale.
		(b) The $+$ exit contrast $\Delta k_+$ from the witness at $t_*$ (points with one standard deviation, $10^4$ shots per setting) and from the projected generator (black curve) on the same scan. Open points flag generators that fail the Markov test. Vertical dotted lines mark the two preference changes, enlarged in the inset.
		(c) Branching pairs at the finite-return Kerr point $(\Delta,G,\widetilde F)/\widetilde U=(1.8,0.5,0.1)$ at $N=60$. The discriminant admits both shaded regions, and the escape test retains the green region. The dot marks the projected Kerr pair. Panels (a) and (b) use the remaining parameters of Fig.~\ref{fig:biased_kerr_directional_branching}.}
	\label{fig:v12bias_checks}
\end{figure*}

\paragraph*{Finite lobe weights.}
At \((\Delta,G,\widetilde F)/\widetilde U=(1.8,0.5,0.1)\), the stationary memberships in the calibrated coordinates are \((\pi_c,\pi_+,\pi_-)=(0.0215,0.0413,0.9372)\). Here \(q/(\Gamma_2\pi_-)=0.398\) and \(p_-(t)\) rises monotonically, while every nonzero exit contrast still peaks at \(\widetilde U t_*=3628\). The reference branching product \(0.381\) exceeds the absorbing-limit value \(0.139\) and lies inside the finite-return feasible region, whose numerical maximum product is \(0.804\) [Fig.~\ref{fig:v12bias_checks}(c)].

In the near-absorbing real-bias scan, the absorbing-center bound and projected branching product increase away from parity on different scales [Fig.~\ref{fig:v12bias_checks}(a)]. At zero bias the individual flip fractions are both about $0.0081$, and the product saturates the small bound. The witness tracks the sign changes of the exit contrast in the same scan [Fig.~\ref{fig:v12bias_checks}(b)]. Throughout, open symbols denote affine generator entries whose interpretation as Markov rates fails.

\subsection{Simulated excitation-map reconstruction and robustness of the witness}
\label{app:rc_measured}

\begin{figure*}[t]
	\centering
	\includegraphics[width=0.857\textwidth]{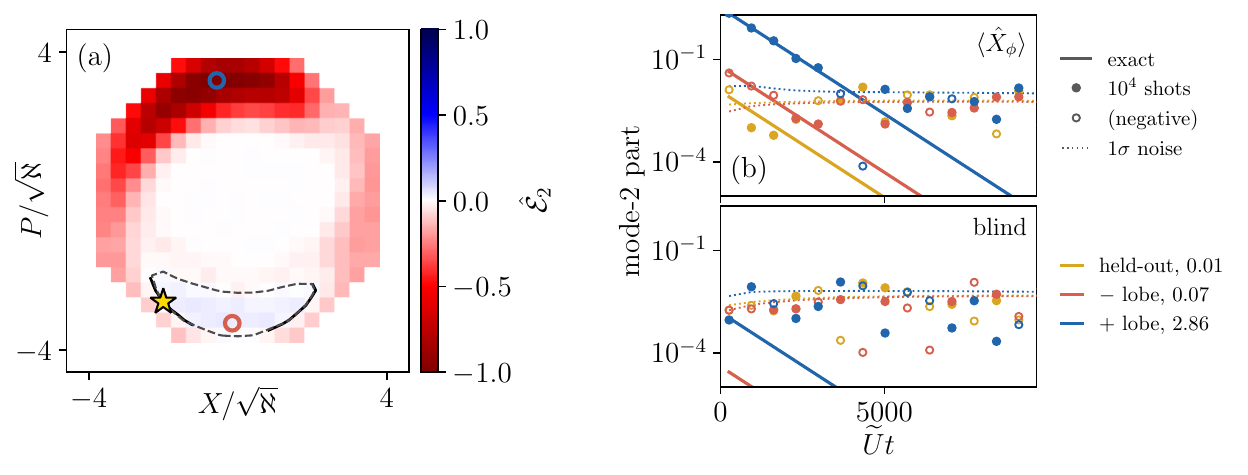}
	\caption{\textit{Excitation map from noisy data and selective preparation at \(\widetilde F/\widetilde U=0.2\).}
		(a) Excitation map of the faster slow mode reconstructed from noisy transients of coherent preparations on a grid of spacing \(0.4\) in \(X/\sqrt\aleph\) and \(P/\sqrt\aleph\), with \(10^4\) shots per preparation, readout, and delay, normalized to its largest modulus. Solid: zero line inferred from the simulated data, drawn only where a neighboring grid point exceeds \(5\%\) of the maximum; dashed: exact zero line. Star: preparation chosen on the inferred zero line; circles: the two lobe preparations.
		(b) Mode-2 part of the calibrated quadrature (top) and of the readout combination blind to mode 2 (bottom) for the three preparations (the preparation on the zero line is labeled ``held-out''), normalized to the largest late signal of each readout; the legend gives the calculated \(|\mathcal E_2/\mathcal E_1|\) in the normalization of Eq.~\eqref{eq:app_mode_amplitude_gauge}. Lines: exact, from full Lindblad propagation. Markers: one simulated record with \(10^4\) shots per readout and delay, the signal minus its fitted constant and mode-1 term (open markers: negative values). Dotted: shot noise (one standard deviation) at a single delay. Points below the dotted lines are dominated by noise; for the blind combination this holds for every preparation.}
	\label{fig:rc_measured}
\end{figure*}

All readout data in this benchmark are simulated; the grid, the delays, and the noise model differ from those of Appendix~\ref{app:v11inference}. For each coherent preparation on the grid, the two readouts at $60$ delays are fitted with the two slow poles of the calibration reconstruction, with Gaussian noise of variance $\operatorname{Var}(\hat O)/10^4$ for each readout and delay. In the exact slow model each nonvanishing modal amplitude matrix over readouts and preparations has rank one; its leading singular component estimates \(\mathcal E_k\) up to normalization, and the calculated maps are used only for comparison. With \(10^4\) shots per preparation, readout, and delay, the reconstructed maps correlate with the exact ones to \(|r|=1.0000\) and \(0.9997\) for the two modes, with relative errors \(0.003\) and \(0.022\); with \(10^3\) shots the mode-2 values are \(0.9967\) and \(0.068\) [Fig.~\ref{fig:rc_measured}(a)]. In the central region, where the excitation amplitude is small, the inferred zero line is not resolved.

A preparation chosen on the inferred near-zero line of \(\mathcal E_2\) and checked by full propagation has \(|\mathcal E_2/\mathcal E_1|=0.013\), compared with \(0.074\) for the \(-\) lobe and \(2.86\) for the \(+\) lobe. Mode 2 carries a fraction \(8.0\times10^{-3}\), \(4.3\times10^{-2}\), and \(0.64\) of the slow amplitude of the quadrature, and \(1.3\times10^{-3}\), \(7.3\times10^{-3}\), and \(0.22\) of that of the photon number. At this bias the \(-\) lobe is already nearly selective (Fig.~\ref{fig:biased_kerr_lr_tomography}); the inferred zero-line preparation reduces this amplitude ratio by a factor of about six in the stated eigenoperator normalization. With \(10^4\) shots per readout and delay, the fitted mode-2 amplitude of the quadrature has the exact sign in all of \(400\) simulated records for both lobes, but it is only marginally resolved for the zero-line preparation: there the median reproduces the exact value, the central \(68\%\) of the records lie between \(0.32\) and \(1.6\) times it, and the sign is correct in \(94\%\) of them [Fig.~\ref{fig:rc_measured}(b)]. A readout combination blind to mode 2 hides the mode for every preparation, with the correct sign in only \(48\%\) to \(59\%\) of the records; this is a property of the readout, not a suppression of the mode.

For the tested calibration and noise model, pairing the quadrature with photon number gives the smaller estimated error for \(\mathcal R_+\): the Fisher-information error of \(\mathcal R_+\) is \(7.50\times10^{-3}\) against \(1.10\times10^{-2}\) for two orthogonal quadratures, and optimizing the quadrature angle lowers it only to \(7.48\times10^{-3}\).

For the witness, a \(1\%\) pole error contributes \(1.6\%\) of the variance at \(\widetilde F/\widetilde U=0.2\) for \(10^4\) shots per setting; it makes the sign unresolvable only where \(|W_+|\) is below twice its pole contribution, whereas the reconstructed column \(Qe_+\) inherits the pole error through both \(h(t)\) and \(g(t)\). The direct and cyclic sign estimates coincide on common data whenever cyclic inversion succeeds. A readout gain error common to calibration and measurement cancels because the coordinates are affine in the readouts; a random gain drift of \(3\%\) between them raises the relative spread of the witness from \(3.2\%\) to \(10\%\) and leaves its sign correct in all realizations.

\section{Matched-pole and destination checks}
\label{app:v11endpoints}

Sparse shift-invert spectra at $N=80,110,140$ use the fixed points $A=(0,5,0.4)$ and $B=(0.45,8.91951842234025,0.409880038986779)$ in $(\theta,\Delta/\widetilde U,G/\widetilde U)$; at these fixed parameters the computed rates at $N=140$ differ by $-4.92\times10^{-9}$ and $4.42\times10^{-10}$ relative. At the separately refitted point $(\Delta,G)/\widetilde U=(8.9195184534374,0.409880038911626)$ the differences fall below $10^{-10}$. Both are below the eigenpair residuals of Table~\ref{tab:endpoint} [residuals $\|(\mathcal L-\lambda_j)r_j\|_2/\|r_j\|_2$ and their adjoint counterparts, divided by $|\lambda_j|$], so they quantify agreement between numerical estimates, not exact equality. The Jacobian of the two log rates with respect to $(\Delta,G)$ is nonsingular at the two finite-difference steps tested. Including complex fast modes, the spectral separation is about $35$.

\begin{table*}[t]
\caption{Endpoint validation for the calibrated transient and the independent detector. Absolute rates and residuals are in units of \(\widetilde U\). The residual ratios use the two decaying modes. The prepared coordinates use the projected frame. The physical readout coordinates use \(M^{-1}\) applied to the stationary means. Stationary lobe coordinates retain the signs of the numerical estimates.}\label{tab:endpoint}
\begin{ruledtabular}\begin{tabular}{lcc}
Diagnostic & \(A\) & \(B\) \\ \hline
Working delay \(t_0\widetilde U\), angle \(\phi\) (rad) & 250.039, 1.634211 & 245.285, 1.634922 \\
\((\pi_+,\pi_-)\), prepared coordinates at \(N=80\) & \(-4.72\times 10^{-16}\), \(3\times 10^{-14}\) & \(7.82\times 10^{-16}\), \(-1.4\times 10^{-15}\) \\
Physical readout \((p_+,p_-)_{\rm ss}\) at \(N=80\) & \(1.36\times10^{-9},\ -1.12\times10^{-9}\) & \(-2.89\times10^{-13},\ 1.35\times10^{-11}\) \\
Noise-whitened calibration condition & 1.754 & 1.411 \\
\(10^5(\Gamma_1,\Gamma_2)/\widetilde U\), \(N=80\) & \(7.4791420189,\ 143.2484017675\) & \(7.4791420191,\ 143.2484017674\) \\
\(10^5(\Gamma_1,\Gamma_2)/\widetilde U\), \(N=110\) & \(7.4791420188,\ 143.2484017674\) & \(7.4791419821,\ 143.2484018310\) \\
\(10^5(\Gamma_1,\Gamma_2)/\widetilde U\), \(N=140\) & \(7.4791420188,\ 143.2484017674\) & \(7.4791419820,\ 143.2484018308\) \\
Max relative slow-pole residual, \(N=140\) & \(2.27\times 10^{-7}\) & \(4.21\times 10^{-7}\) \\
Slow-mode condition numbers, \(N=140\) & 6.96, 2.53 & 4.97, 2.68 \\
Log-rate Jacobian \(\sigma_{\min}\), \(N=140\) & 0.218 & 0.118 \\
Maximum detector identity-closure error & \(1.31\times 10^{-7}\) & \(7.52\times 10^{-8}\) \\
Maximum absolute detector adjoint residual, \(N=140\) & \(1.36\times 10^{-9}\) & \(4.11\times 10^{-9}\) \\
Least eigenvalue of numerical \(\hat F_j\), \(N=140\) & \(7.49\times 10^{-11}\) & \(-6.78\times 10^{-11}\) \\
Maximum physical embedding residual \(\|\mathcal D_j\|_1/\widetilde U\) & \(1.57\times 10^{-7}\) & \(1.02\times 10^{-7}\) \\
Dual defect, fixed input frame at \(t_0\) & \(2.12\times 10^{-12}\) & \(1.21\times 10^{-11}\) \\
Minimum off-diagonal \(Q_{ji}/\widetilde U\) (unresolved) & \(-2.72\times 10^{-18}\) & \(-2.84\times 10^{-19}\) \\
\(10^4\Delta k_+/\widetilde U\), representative and jitter range & \([4.409,\ 4.572]\) & \([-6.090,\ -6.015]\) \\
Held-out maximum coordinate error & \(1.18\times 10^{-9}\) & \(1.4\times 10^{-11}\) \\
Estimated blind readout residual detection factor & \(3.12\times 10^{-12}\) & \(1.41\times 10^{-11}\) \\
Estimated blind readout held-out max. absolute error & \(3.74\times 10^{-9}\) & \(3.38\times 10^{-10}\) \\
\end{tabular}\end{ruledtabular}
\end{table*}

The effects in Eq.~\eqref{eq:v11effects} are compressions of the coherent-state resolution of identity. The disks are disjoint, and the compressed kets retain their projected norms. In a finite Fock space every open disk gives a positive-definite effect because a coherent-overlap polynomial that vanishes on the disk vanishes identically. For positive strengths the surviving trace therefore vanishes asymptotically, since
\begin{equation}
	\frac{\mathrm d}{\mathrm dt}\operatorname{Tr}(\hat\rho_{\rm surv})
	\le -\lambda_{\min}\left(\sum_j\gamma_j\hat E_j\right)\operatorname{Tr}(\hat\rho_{\rm surv}).
\end{equation}
For $\hat E_j=\sum_l e_{jl}|v_{jl}\rangle\langle v_{jl}|$, the jumps $\sqrt{\gamma_j e_{jl}}\,|j_{\rm sink}\rangle\langle v_{jl}|$ give a completely positive realization in an enlarged space. A device would need couplings that realize these jump effects and separate monitoring of the two sinks. In exact arithmetic $\hat F_j$ is positive and $\sum_j\hat F_j=\hat I$; we solve the two adjoint linear equations independently, check their residuals, positivity, and completeness, and compare with forward integration at small cutoff. The disk integrals use Gauss--Legendre nodes in squared radius and uniform angles ($20\times80$, checked against $32\times128$). Across $N=80,110,140$ and the sampled strengths, radii, and input shifts $\delta\alpha=\pm0.12,\pm0.12i$, the detection probabilities change with the cutoff by at most $4.41\times10^{-7}$.

For equal strengths $\gamma_c=\gamma_-=\gamma$, write $\hat E=\hat E_c+\hat E_-$ and $\mathcal K_\gamma=\mathcal L-\gamma\{\hat E,\,\cdot\,\}/2$. If the finite-dimensional unmonitored Liouvillian is relaxing, its isolated stationary eigenvalue shifts to $-\gamma\operatorname{Tr}(\hat E\hat\rho_{\rm ss})+O(\gamma^2)$, with spectral projector tending to $\hat\rho_{\rm ss}\operatorname{Tr}(\,\cdot\,)$. The other modes give a vanishing contribution to $\gamma\int_0^\infty\operatorname{Tr}[\hat E_-e^{t\mathcal K_\gamma}\hat\rho_+]\,\mathrm dt$ as $\gamma\to0$. For nonzero stationary detector weight this yields
\begin{equation*}
	\lim_{\gamma\to0}\mathcal P_-=
	\frac{\operatorname{Tr}(\hat E_-\hat\rho_{\rm ss})}
	{\operatorname{Tr}[(\hat E_c+\hat E_-)\hat\rho_{\rm ss}]}
	\;.
\end{equation*}
The limit is independent of the normalized initial preparation at fixed finite cutoff and fixed relaxing Liouvillian. It does not assert uniform convergence if the unmonitored spectral gap also tends to zero.

At half the damping, $\kappa/\widetilde U=0.05$, a separate pair matches rates $2.50\times10^{-5}\widetilde U$ and $4.92\times10^{-4}\widetilde U$. At the reference detector settings, both favor the opposite lobe with probabilities $0.849$ and $0.645$. At $r=0.75$ and $\gamma/\widetilde U=10^{-4}$ the probabilities are $0.529$ and $0.403$ and the signs survive the tested input shifts at $N=140$. The window of detector strengths that resolves the reversal therefore shifts with the damping. These checks concern the single-mode loss model. The main operating points have $\kappa/K=1$, with $K=U/2$, above the damping ratios typical of the Kerr-cat devices of Refs.~\cite{Grimm2020,Frattini2024,CrosCarrilloDeAlbornoz2024}.

\section{Physical closure of the readout reduction}
\label{app:v11closure}

For a trace-one Hermitian representative $\hat\sigma_j$, the negative spectral weight $(\|\hat\sigma_j\|_1-1)/2$ equals its smallest trace distance to a physical state. A measurement of the projector onto the negative space gives a lower bound and normalizing the positive part attains it. At $A$ and $B$ ($N=80$), the left-accessible representatives used for $\mathbf Q$ have maximum negative weight $0.052$ and $0.074$ (Fig.~\ref{fig:biased_kerr_directional_branching} quotes instead $\varepsilon_\sigma$, the magnitude of their most negative eigenvalue), the slow projections of the coherent anchors at preparation $0.014$ and $0.030$, the same projections at the plateau time $t_0$, which define the prepared basis, at most $4\times10^{-16}$, and the alternative representatives inscribed in the physical states of the slow slice are positive to $10^{-12}$. For coordinate duals $\hat\eta_j$, the exact all-input coordinate defect is
\begin{equation}
	\max_{\varepsilon_j=\pm1}\lambda_{\max}\!\left(\sum_j\varepsilon_j\hat\eta_j\right)-1 .
\end{equation}
This is twice the largest total negative coordinate weight~\cite{Macieszczak2021}, since $\sum_jp_j=1$ implies $\sum_j|p_j|-1=2\sum_{j:p_j<0}|p_j|$. It is $0.084$ and $0.090$ in the prepared frames over all density matrices and $0.0062$ and $0.0039$ on the sampled coherent grids. Changing anchors changes $Q$ by similarity; the signed $+$ contrast keeps its sign when each anchor is displaced in turn or when positive representatives are used (Table~\ref{tab:endpoint}). The detector probabilities are computed without this reduction.

Let $\mathsf W$ have as columns the three physical density matrices at $t_0$ and put $\mathcal D=\mathcal L\mathsf W-\mathsf WQ$. The identity
\begin{equation}
	e^{t\mathcal L}\mathsf W-\mathsf W e^{tQ}
	=\int_0^t e^{(t-s)\mathcal L}\mathcal D e^{sQ}\,\mathrm ds
\end{equation}
relates the embedding residual to the propagated error for an input represented by $\mathsf Wp(0)$; a general input adds its initial embedding error $\|\hat\rho(0)-\mathsf Wp(0)\|_1$, which does not grow by contractivity. For a Markov generator $Q$ and a probability vector $p(0)$, the error is bounded by $t\max_j\|\mathcal D_j\|_1$. We test the reduction on six preparations per operating point: five held-out states, namely three coherent states displaced by $0.4$ in scaled units from the lobe attractors, the even cat of the two lobes, and a displaced thermal state with $\bar n_{\rm th}=0.3$, and, as a linearity check, the equal mixture of the two lobe anchors. Each is prepared at $t=0$ and evolved with the full Liouvillian to the plateau origin $t_0$, after the fast transient. From their quadrature and photon-number means at $t_0$, the calibrated generator and embedding $\mathsf W$ predict the density matrices and independent observables at delays up to $10t_*$ after $t_0$: the orthogonal quadrature, $\hat n^2$, parity, and a combination of $\hat X_\phi$ and $\hat n$ fitted to suppress the faster mode on the $+$ trace. The maximum trace distances are $1.44\times10^{-8}$ at $A$ and $5.55\times10^{-9}$ at $B$ on the sampled delays. These values quantify the accuracy of the slow-sector description after the fast transient, not tomography of arbitrary states from two means. The maximum calibrated-coordinate errors and the residual detection and prediction errors of the fitted blind combination are given in Table~\ref{tab:endpoint}.

\section{Information on an eigenline and finite-shot acquisition}
\label{app:v11inference}

When $v_0\ne0$ lies on an eigenline $L$, choose a vector $w_2$ completing $v_0$ to a basis of the plane. With the observed eigenvalue $-\Gamma_i$ and the other pole $-\Gamma_j$, the full ambiguity on the plane is the arbitrary upper-right entry of the triangular matrix with these diagonal entries. Extended to the three coordinates, the ambiguity is $\delta Q=\zeta v_0 f^{\mathsf T}$ where $f^{\mathsf T}v_0=f^{\mathsf T}\pi=0$ and $f^{\mathsf T}w_2=1$. Consequently $d^{\mathsf T}Qw$ is fixed exactly when $(d^{\mathsf T}v_0)(f^{\mathsf T}w)=0$ in this affine family. For a single-mode lobe anchor, $w=e_\pm=\pi+v_0$ gives $f^{\mathsf T}w=0$: its full column, including both exit entries, is identifiable while the action along $w_2$ remains free. A readout with zero detection of one mode is a separate obstruction. Two calibrated readouts that are linearly independent on the deviation plane remove it, but no readout restores a mode absent from the prepared state.

\paragraph*{Several preparations.}
With independently known poles and stationary coordinates, Eq.~\eqref{eq:rc_kernel} can be inverted at one delay for several preparations at once. If the columns of $P_0$ and $P(t)$ hold the calibrated coordinates of the preparations at $t_0$ and after a delay $t$, and their deviations $V_0=P_0-\pi\mathbf 1^{\mathsf T}$ span the zero-mass plane,
\begin{equation}
	Q=\frac{P(t)-\pi\mathbf 1^{\mathsf T}-h(t)V_0}{g(t)}\,V_0^{+}(I-\Pi)
	\;,
	\label{eq:rc_one_time}
\end{equation}
where $V_0^+$ is the Moore--Penrose pseudoinverse; this uses $e^{tQ}v=h(t)v+g(t)Qv$ on the plane and $Q\Pi=0$. Two preparations with independent deviations suffice; for the three anchors $P_0=I$ in the prepared basis. At $\widetilde F/\widetilde U=0.2$, $150$ simulated reconstructions with $10^5$ shots per preparation, readout, and delay give $B_{+\to-}=0.661\pm0.001$ and $B_{-\to+}=0.450\pm0.005$.

\paragraph*{Repeated, complex, and additional slow poles.}
In the degenerate limit \(\Gamma_1,\Gamma_2\to\Gamma\), \(g(t)\to te^{-\Gamma t}\), \(h(t)\to(1+\Gamma t)e^{-\Gamma t}\), and \(t_*\to1/\Gamma\); Eq.~\eqref{eq:rc_kernel} then also holds for a generator with a Jordan block, while a diagonalizable one equals \(-\Gamma(I-\Pi)\) and is fixed by \(\Gamma\) and \(\pi\). With an absorbing center, Eq.~\eqref{eq:rc_absorbing_bound} then forbids any round trip, \(B_+B_-=0\). For a repeated real pole with a Jordan block, $v_0$ and $Jv_0$ span the plane whenever $Jv_0\neq0$, and the Cayley--Hamilton reconstruction holds with $S=2\Gamma$, $D=\Gamma^2$, and the limiting kernel. For conjugate poles $-\Gamma\pm i\omega$ the kernel becomes $e^{-\Gamma t}\sin(\omega t)/\omega$, and at its zeros the delayed action cannot be recovered, although every real nonzero deviation is then cyclic. In higher dimension, a single delay generally does not supply enough independent Krylov directions; the one-delay extraction relies on the exponential reducing to a linear polynomial on the two-dimensional plane. With $m\geq2$ distinct real nonstationary slow poles and full calibrated coordinates, the deviations at $m$ distinct times $0,t_1,\dots,t_{m-1}$ form $V=R\operatorname{diag}(\mathcal E_1,\dots,\mathcal E_m)T$, where $T_{kj}=e^{\lambda_kt_j}$ is a known nonsingular Vandermonde-type matrix. Hence $R\operatorname{diag}(\mathcal E)=VT^{-1}$, and $Q$ on the deviation space equals $VT^{-1}\Lambda(VT^{-1})^{-1}$. One preparation and $m-1$ positive delays therefore suffice exactly when $\prod_{k=1}^m\mathcal E_k\neq0$, the cyclic-vector criterion of Eq.~\eqref{eq:v11det} for $m$ modes.

\begin{figure}[!b]
	\centering
	\includegraphics[width=0.96\linewidth]{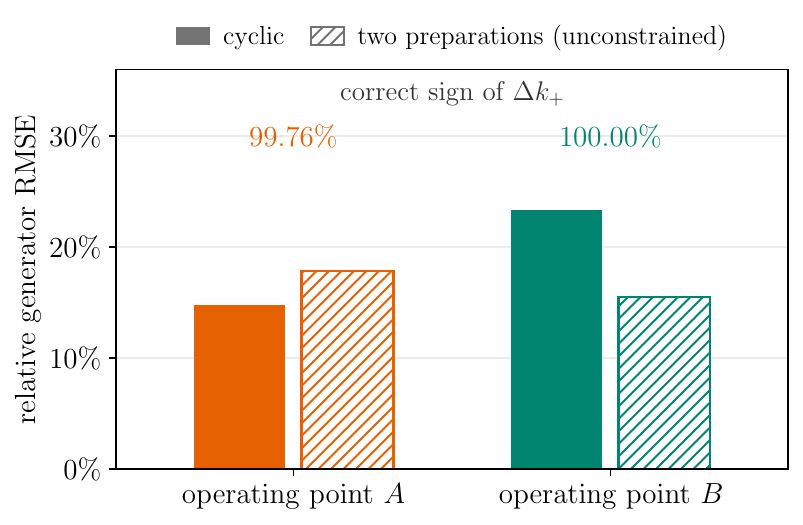}
	\caption{\textit{Directional sign and full-generator recovery with calibration, drift, and pole uncertainty.} Results at $80$ delayed shots per readout channel over $5000$ trials per operating point; this budget counts the delayed acquisition only, not the calibration or the pole estimation. Solid and hatched bars show the full-generator relative Frobenius RMSE for cyclic reconstruction and for the unconstrained two-preparation formula, Eq.~\eqref{eq:rc_one_time}, respectively. The percentages above the bars give the direct estimate's accuracy for the sign of $\Delta k_+$.}
	\label{fig:v16_endpoint_benchmark}
\end{figure}

\paragraph*{Matched operating-point benchmark.}
Figure~\ref{fig:v16_endpoint_benchmark} samples the exact finite-Fock distributions of $\hat X_\phi$ and $\hat n$ after full Lindblad evolution at $A$ and $B$. In each trial, all estimators use the same sampled poles and the same calibration estimated from finite-shot data ($10^3$ shots per channel for each plateau anchor and $2000$ for the stationary readout). The model of calibration and pole uncertainties adds correlated gain perturbations of scale $0.004$, offsets of $0.002$ times the channel calibration span, and correlated log-pole errors of order $1\%$, with correlations $0.4$ across channels, $0.6$ between calibration and delayed acquisition, and $0.2$ across both. At $80$ delayed shots per channel, the direct sign estimate succeeds in $99.76\%$ and $100\%$ of $5000$ trials at $A$ and $B$ ($95\%$ intervals $[99.58,99.88]\%$ and $[99.93,100]\%$), a two-time fit at $0.83t_*$ and $1.25t_*$ in $99.90\%$ and $100\%$, and the unconstrained two-preparation formula in $97.84\%$ and $99.78\%$. The cyclic full-generator errors are $14.7\%$ and $23.4\%$, almost unchanged ($14.6\%$, $22.9\%$) with these uncertainties held fixed; no inversion abstains at this budget. The cost estimate excludes the waiting time for stationarity and the estimation of the poles.

\begin{figure}[t]
	\centering
	\includegraphics[width=0.88\linewidth]{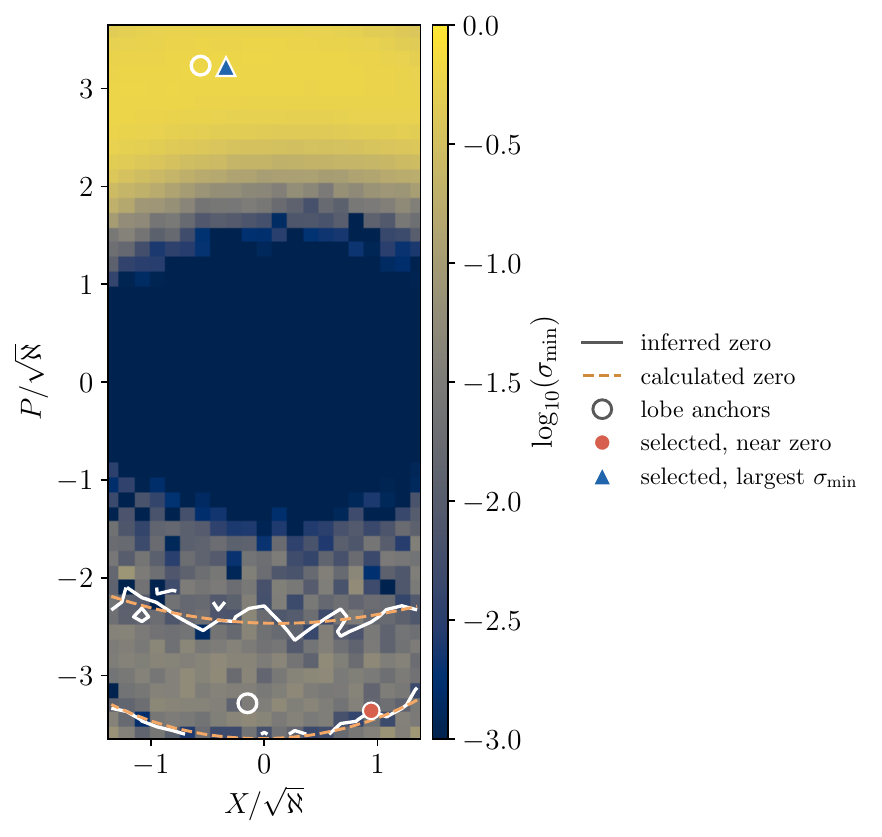}
	\caption{\textit{Conditioning inferred from simulated transients at $A$.} Color: $\log_{10}(\sigma_{\min})$ of the scaled cyclic basis, inferred from noisy quadrature and photon-number transients of $21\times49$ coherent preparations ($4096$ shots per setting, six delays). The dark central region reflects preparations that are nearly stationary after $t_0$ ($v_0\simeq0$), not a selective suppression of the faster mode. White: inferred zero of the faster-mode amplitude; dashed orange: calculated zero; both are drawn in the negative-lobe region $P/\sqrt\aleph<-2$ where the slower-mode amplitude exceeds half its maximum. Open circles: lobe anchors; filled circle: input selected on the inferred zero ($\sigma_{\min}=0.017$); triangle: input with the largest inferred $\sigma_{\min}$ ($0.619$).}
	\label{fig:pilot_map}
\end{figure}

\paragraph*{Learning the preparation from pilot data.}
At $A$ and $N=80$, the pilot uses $21\times49$ coherent inputs and six delays $\{0,0.12,0.33,0.68,1.40,3.00\}t_*$ after $t_0$, with Gaussian sample means whose variances follow from the finite-Fock outcome distributions at $4096$ shots per channel ($5.1\times10^7$ repetitions in total); poles and stationary readouts are fixed, and the measured operators are evolved adjointly by rational Arnoldi. On each input we subtract the stationary means, fit the two known exponentials, take the leading rank-one factor per pole, and convert the readout-space modal contributions to coordinates through $M^{-1}$ (Fig.~\ref{fig:pilot_map}). One input is selected on the fitted faster-mode zero in the negative-lobe region, another at the largest forecast $\sigma_{\min}$ ($0.618$; independent forward evolution gives $0.619$). Over $64$ redraws of the pilot, the 5th--95th percentile bands of the actual $\sigma_{\min}$ are $[0.013,0.032]$ and $[0.617,0.619]$; at $512$ shots per setting they are $[0.031,0.041]$ and $[0.613,0.617]$.

The selected inputs are evolved independently (two rational approximations agree within $1.2\times10^{-7}$ in Hilbert--Schmidt norm, and an $N=110$ check changes the calibrated coordinates by less than $8\times10^{-12}$). Every validation trial draws fresh multinomial samples for the delayed and initial means, and the cyclic inversion is accepted only if $\sigma_{\min}>10^{-5}$ and $\sigma_{\max}/\sigma_{\min}<500$; rejections count as failures. At the exact excitation zero the noiseless basis is rank deficient, and any apparent invertibility under noise is spurious. With $4096$ initial shots and $80$, $320$, or $1280$ delayed shots per channel, the fraction of trials with relative Frobenius error below $25\%$ is $16$, $20$, and $23\%$ for the input selected on the inferred zero and $20$, $28$, and $49\%$ for the $-$ anchor. It is $93$, $100$, and $100\%$ both for the input with the largest inferred $\sigma_{\min}$ and for the $+$ anchor. We then add the input with the largest $\sigma_{\min}$ to the one selected on the inferred zero, split both the initial and the delayed budgets equally between them, and use the unconstrained formula of Eq.~\eqref{eq:rc_one_time}. The fraction at $80$ delayed shots rises from $16.3\%$ to $88.7\%$, and to $100\%$ at $320$ and $1280$. The second preparation probes the direction that the nearly selective one misses, but at equal budget the pair does not outperform the well-conditioned input alone ($93.3\%$). With the supplied calibrations, the pilot data alone already suffice to estimate $Q$; the pilot shows that the conditioning can be inferred from transients, not that it lowers the total acquisition cost.

\paragraph*{Shot scaling.}
Figure~\ref{fig:v11identifiability}(c) uses the exact finite-Fock outcome distributions of $\hat X_\phi$ and $\hat n$ at $t_0$ and $t_*$ after full Lindblad evolution of the two lobe anchors at $A$ ($N=80$), stored with the validation run above. For every budget $N_{\rm shots}=2^5,\dots,2^{17}$ and preparation, $4000$ trials draw $N_{\rm shots}$ multinomial outcomes per channel and time; poles and calibration are fixed to their exact values, the rank gate is the one above, and abstentions count as infinite error. The two-anchor estimate uses $N_{\rm shots}/2$ outcomes per preparation, channel, and time. Like the cyclic estimator, which enforces the poles through the Cayley--Hamilton relation, it imposes the known poles. In an orthonormal basis of the zero-mass plane, the restriction of $Q$ is parametrized as $\big(\begin{smallmatrix}a & b\\ c & -S-a\end{smallmatrix}\big)$ with $c=[a(-S-a)-D]/b$, so that its trace is $-S$ and its determinant $D$ for all $(a,b)$ with $b\neq0$. At $A$ and $B$ the exact restrictions have $|b|/\|Q_\perp\|=0.242$ and $0.249$, far from the singular chart $b=0$. The two parameters minimize the unweighted sum of squared residuals $\|Q_\perp V-A\|_F^2$, where the columns of $V$ and $A$ are the two measured deviations $v_0$ and actions $Qv_0$ in that basis. The fit uses a trust-region least-squares solver (SciPy \texttt{least\_squares}) started from the unconstrained solution, with tolerances $10^{-15}$; the restriction to the plane is then mapped back with $Q\pi=0$. A fit that returned no finite estimate would count as a failure; none occurred in the $4000$ trials at any budget, including $N_{\rm shots}=32$. The asymptotes $c/\sqrt{N_{\rm shots}}$ are the medians of $\|J\xi\|_F/\|Q\|_F$ for Gaussian $\xi$ with the single-shot covariances of the measured means, with finite-difference Jacobians $J$ of the estimators; they give $c=9.32$, $1.08$, and $1.03$ for the $-$ anchor, the $+$ anchor, and both anchors, and the simulated medians follow them at large $N_{\rm shots}$. On the sampled grid the simulated medians reach $10\%$ at $N_{\rm shots}=128$ for the $+$ anchor and $8192$ for the $-$ anchor, a factor $64$, compared with $(9.32/1.08)^2\simeq75$ from the asymptotes. Two estimator-level checks use the same outcome distributions. First, if the anchor coordinates at $t_0$ are supplied exactly ($v_0=e_\pm-\pi$, as the calibration defines them) and only the delayed outcomes are sampled, the prefactors become $8.60$ and $1.00$, a ratio of $74$; the simulated medians at $N_{\rm shots}=8192$ are $0.095$ and $0.011$. The cost ratio is therefore set by the information in the delayed data, not by the treatment of the initial coordinates. Second, the unconstrained two-preparation formula, Eq.~\eqref{eq:rc_one_time}, whose estimate need not have the prescribed poles, gives $c=1.53$ instead of $1.03$. The apparent penalty for splitting the budget between the two anchors is thus a consequence of the missing pole constraint, and comparisons between preparation strategies are meaningful only for equally constrained estimators.

\paragraph*{Information bound.}
With $\pi$, the poles, and the initial coordinates fixed, the two local parameter directions are the shears of Eq.~\eqref{eq:isospectral_sensitivity}, which preserve the poles to first order. For readout means $y=M_2p$ at a delay $t$, with $M_2$ the first two rows of the calibration and per-shot variances $\Sigma=\operatorname{diag}[\operatorname{Var}(\hat X_\phi),\operatorname{Var}(\hat n)]$ from the exact outcome distributions, the asymptotic Fisher matrix per shot of the sample means is $I=J^{\mathsf T}\Sigma^{-1}J$ with $J=g(t)[\mathcal E_2M_2r_1,\ \mathcal E_1M_2r_2]$. Propagating the asymptotic Cram\'er--Rao covariance $I^{-1}/N_{\rm shots}$ to $\|\delta Q\|_F/\|Q\|_F$ at $t_*$ gives median-error prefactors $8.61$ and $1.003$ for the $-$ and $+$ anchors, equal to the prefactors $8.60$ and $1.00$ of the cyclic estimator with known initial coordinates, and a ratio of $73.6$. On the pilot grid, the exact means at the six delays give the modal amplitudes of each preparation. We evaluate $I$ at the delay $1.40t_*$ for the $219$ preparations with $P/\sqrt\aleph<-2$, $|\mathcal E_1|$ above half of its maximum in that region, and $|\mathcal E_2|$ below $20\%$ of its maximum (spanning $4\times10^{-4}$ to $0.1$); a power-law fit gives $\lambda_{\min}(I)\propto|\mathcal E_2|^{2.07}$, consistent with the quadratic loss predicted by Eq.~\eqref{eq:isospectral_sensitivity}. For the distance law, the zero contour is located by linear interpolation of $\mathcal E_2$ between neighboring grid points along $P$, and $d_\perp$ is the distance of each preparation to the nearest interpolated zero; the linearized estimate $|\mathcal E_2|/|\nabla\mathcal E_2|$ is a distance only close to the contour. For the $49$ preparations with $d_\perp<0.1\sqrt\aleph$, the Schur complement of Eq.~\eqref{eq:schur_information} scales as $d_\perp^{1.94}$ and as $|\mathcal E_2|^{2.03}$; including preparations up to $0.5\sqrt\aleph$ lowers the first exponent to $1.79$, consistent with the nonlinear variation of $\mathcal E_2$ away from the contour. At a higher-order zero, $\mathcal E_2\propto s^\nu$ along a path parametrized by $s$, the same argument gives a shot number growing as $|s|^{-2\nu}$. The eigenline relation for $\sigma_{\min}$ in Sec.~\ref{subsec:measurement_cost} follows from $|\det(K)|=(\Gamma_2-\Gamma_1)|\mathcal E_1\mathcal E_2|\,\|r_1\|\|r_2\||\sin(\vartheta)|/S$ and $\sigma_{\max}\simeq|\mathcal E_1|\,\|r_1\|(1+\Gamma_1^2/S^2)^{1/2}$ for small $\mathcal E_2$; for the $-$ anchor it gives $0.0401$ against the computed $0.0398$, while for the $+$ anchor, far from the eigenline, it does not apply.

\FloatBarrier
\bibliographystyle{KilianStyle_v2}
\bibliography{biblio_L_tomography_v13}

\end{document}